\documentclass[iop,superscriptaddress]{emulateapj}

\usepackage{graphicx}
\usepackage{dcolumn}
\usepackage{amssymb} 
\usepackage{ctable}
\usepackage{bm}
\usepackage{epsfig}
\usepackage{epstopdf}
\usepackage{epsf,color}
\usepackage{natbib}

\newcommand{\cmnt}[1]{}

\def\vecx{\mathbf{x}}

\def\veck{\mathbf{k}}
\def\hatn{\mathbf{\hat n}}

\def\VEV#1{\left\langle #1 \right\rangle}

\begin{document}

\title{Intensity Mapping across Cosmic Times with the Lyman-Alpha Line}

\author{Anthony R. Pullen}
\affiliation{NASA Jet Propulsion Laboratory, California Institute of Technology, 4800 Oak Grove Drive, MS 169-237, Pasadena, CA, 91109, U.S.A.}
\author{Olivier Dor\'{e} and Jamie Bock}
\affiliation{NASA Jet Propulsion Laboratory, California Institute of Technology, 4800 Oak Grove Drive, MS 169-327, Pasadena, CA, 91109, U.S.A.}
\affiliation{California Institute of Technology, MC 249-17, Pasadena, California, 91125 U.S.A.}

\email{anthony.r.pullen@jpl.nasa.gov}

\begin{abstract}
We present a quantitative model of Lyman-alpha (Ly$\alpha$) emission throughout cosmic history and determine the prospects for intensity mapping of spatial fluctuations in the Ly$\alpha$ signal.  Since i) our model assumes at $z>6$ the minimum star formation required to sustain reionization and ii) is based at $z<6$ on a luminosity function extrapolated from the few observed bright Ly$\alpha$ emitters, this should be considered a lower limit.  Mapping the line emission allows probes of reionization, star formation, and large-scale structure (LSS) as a function of redshift.  While Ly$\alpha$ emission during reionization has been studied, we also predict the post-reionization signal to test predictions of the intensity and motivate future intensity mapping probes of reionization.  We include emission from massive dark matter halos and the intergalactic medium (IGM) in our model.  We find agreement with current, measured luminosity functions of Ly$\alpha$ emitters at $z<8$.  However, diffuse IGM emission, not associated with Ly$\alpha$ emitters, dominates the intensity up to $z\sim10$. While our model is applicable for  deep-optical or near-infrared observers like the SuMIre survey\footnote{\citet{2012arXiv1206.0737E}} or the James Webb Space Telescope, only intensity mapping will detect the diffuse IGM emission. We also construct a 3D power spectrum model of the Ly$\alpha$ emission, and characterize possible foreground contamination.  Finally, we study the prospects of an intensity mapper for measuring Ly$\alpha$ fluctuations while identifying interloper contamination for removal.  Our results suggest that while the reionization signal is challenging, Ly$\alpha$ fluctuations can be an interesting new probe of LSS at late times when used in conjunction with other lines like, e.g., H$\alpha$, to monitor low-redshift foreground confusion.
\end{abstract}
\maketitle
 
\section{Introduction} \label{S:intro}
Much effort has been spent attempting to elucidate the Epoch of Reionization (EoR) \citep{2001PhR...349..125B}, when the matter in the universe, dominated by hydrogen gas, transformed from the neutral state to an ionized state.  Considering this phase transition is most likely a product of the first population of stars, probing this epoch opens a window into star formation and large-scale structure (LSS) in the early universe.  Many questions concerning this epoch remain unanswered, including the nature of the first stars and when reionization occurred, although evidence from the \textit{Wilkinson Microwave Anisotropy Probe} (WMAP) \citep{2012arXiv1212.5225B} has suggested that reionization began around redshift $z\simeq11$ and was completed around $z\simeq7$ \citep{2012arXiv1212.5226H}.  This picture is also consistent with the recently released Planck observations \citep{2013arXiv1303.5062P,2013arXiv1303.5076P}.  We also see further evidence of Ly$\alpha$ absorption by neutral hydrogen at high redshifts from the detection of Gunn-Peterson troughs in quasar spectra \citep{Fan:2004bn} and damped Ly$\alpha$ systems \citep{Totani:2005ng}.  The characterization of fluctuations in the near-infrared background (NIRB) due to continuum emission of early stars during the reionization epoch has been attempted \citep{2005Natur.438...45K,2010ApJ...710.1089F,2011ApJ...742..124M,2012ApJ...750...20F,2012ApJ...756...92C}.  The fluctuations cominate over low-redshift galaxies and appear to exhibit a Rayleigh-Jeans spectrum \citep{2011ApJ...742..124M}.  \citet{2005Natur.438...45K} claims the fluctuations are from the EoR while \citet{2012Natur.490..514C} claims the fluctuations are from diffuse intrahalo light (IHL).  Both possibilities would have significant implications for future Ly$\alpha$ studies, especially since IHL would be a serious foreground to Ly$\alpha$ clustering.  In addition, continuum emission cannot be used to map the tomography of reionization sources because emission at various redshifts combine in intensity maps in each frequency band.

To answer questions regarding the nature of the EoR and star formation, line emission provides essential redshift information.  Intensity mapping has arisen as a potentially powerful probe of LSS and star formation at high redshifts \citep{1999ApJ...512..547S,2008A&A...489..489R,2010JCAP...11..016V}.  Unlike spectroscopic surveys, which use catalogs of objects detected individually, an intensity map captures the aggregated light from an emission line in large pixels across the sky, including emission from unresolved sources.

Various lines have been proposed as potential candidates for intensity mapping, including HI \citep{2011ApJ...728L..46G}, CO  \citep{2008A&A...489..489R,2011ApJ...730L..30C,2011ApJ...728L..46G,2011ApJ...741...70L,2013ApJ...768...15P,2013MNRAS.tmp.2130M}, CII \citep{2012ApJ...745...49G}, and Ly$\alpha$ \citep{2013ApJ...763..132S}.  In this study we focus on Ly$\alpha$ emission.  Ly$\alpha$ photons emitted during the EoR with rest-frame wavelength 1216\AA\, will redshift to the near-infrared regime today, making them potentially detectable by narrow-band infrared detectors.  Zodiacal light emission makes detecting the absolute photometry in the infrared problematic; however, we can circumvent this issue by mapping the spatial fluctuations in the Ly$\alpha$ line emission on small scales where the zodiacal light is spatially smooth.  Previous work by \citet{2013ApJ...763..132S} described the Ly$\alpha$ intensity mapping signal for during the EoR, specifically redshifts $z=[7,11]$.  We extend their work by filling in the post-reionization epoch, all the way down to $z=2$, considering that many telescopes [e.g. Hyper Suprime Cam (HSC), the James Webb Space Telescope (JWST)], may seek to measure the low-redshift signal in order to constrain cosmological parameters as well as star formation.

Ly$\alpha$ emission in the EoR and post-EoR comes from both massive, dark-matter halos containing ionizing stellar populations and from the surrounding IGM.  The emission in the halos is predicted to be sourced mainly by Pop-III and Pop-II stars that emit ionizing photons which ionize neutral hydrogen, which then emits Ly$\alpha$ photons upon recombination.  These stellar populations can also source emission in the IGM through $e$-$p$ recombinations and $e$-HI collisions.  In addition, diffuse IGM emission should be sourced by $e$-$p$ recombinations in the IGM, as well as photons from continuum emission from higher-redshift stars, which are then redshifted, absorbed, and re-emitted at Ly$\alpha$ photons.  The relative contributions of the halos and stellar IGM are partly determined by $f_*$, the star formation efficiency, $f_{\rm esc}$, the escape fraction of ionizing photons from the halos to the IGM, the gas temperature, and $M_{\rm min}$, the minimum mass of Ly$\alpha$ halos, while the diffuse IGM emission from continuum stars depends on $f_*$, the stellar spectrum, and the ionization history.

In this paper, we model the theoretical intensity signal of the Ly$\alpha$ line over cosmic history, specifically over redshifts $z=2$ to 12.  We also construct an empirical estimate of the Ly$\alpha$ intensity out to $z=8$ based on measurements of the Ly$\alpha$ luminosity function at various redshifts.  We find that although the empirical estimate is slightly lower than theoretical estimates from Ly$\alpha$ emitters for the intensity, these estimates do agree to within an order of magnitude.  We find that for Ly$\alpha$ emitters, the emission from massive halos dominates the signal at all redshifts considered ($z<12$).  However, the emission components from the diffuse IGM dominate the intensity over all the other sources, except at high redshifts ($z\gtrsim10$) where Ly$\alpha$ emitters dominate again.

Using these estimates of the intensity, we calculate the 3D power spectrum of Ly$\alpha$ line emission from fluctuations in the halos and the IGM, implementing a halo-model formalism for the clustering bias.  As with the intensity, diffuse IGM emission dominates the signal for most of our redshift range, meaning cross-correlations with other emission lines will be required to detect the Ly$\alpha$-emitter signal.  Note that these estimates do not include fluctuations in the ionization fraction, which may increase the signal for all the IGM-related sources above our estimates.

We also study prospects for measuring the Ly$\alpha$ power spectrum.  We consider an intensity mapper with reasonable specifications for measuring Ly$\alpha$ fluctuations over the redshift range $3<z<12$ for a 200 deg$^2$ survey, and find that Ly$\alpha$ emission should be detectable for tomography at late and post-reionization epochs ($z_{\rm Ly\alpha}=3-8$), but fluctuations during earlier times during reionization are too low in our fiducial model for detection, except possibly for an extremely deep survey. We then estimate foreground emission that will contaminate the signal.  We find that line emission foregrounds such as H$\alpha$ emission lines that contaminate Ly$\alpha$ maps from redshifts $z=3-6$ can be masked to reveal the Ly$\alpha$ power spectrum, but this gets increasingly difficult for higher redshifts.

The plan of our paper is as follows:  Sec.~\ref{S:emp} constructs an empirical estimate of the Ly$\alpha$ intensity out to $z=8$ based on luminosity function measurements. In Sec.~\ref{S:modint}, we model the theoretical intensity signal of the Ly$\alpha$ line out to $z=12$.  We construct the theory for the fluctuations in the Ly$\alpha$ signal, including for the IGM, in Sec.~\ref{S:modfluc}.  We present forecasts in Sec.~\ref{S:forecasts}, and present our conclusions in Sec.~\ref{S:conc}.  Wherever not explicitly mentioned, we assume a flat $\Lambda$CDM cosmology with parameters compatible with WMAP7.

\section{Minimum Lyman-alpha emission - Empirical}\label{S:emp}
We begin by constructing an empirical estimate of the Ly$\alpha$ intensity based on luminosity function (LF) measurements of Ly$\alpha$ emitters out to redshift $z\sim8$.  This estimate is essentially a lower limit in that LFs are based on a few bright detections and are extrapolated to lower luminosities.  The standard relation for intensity from cosmological distances is given by
\begin{eqnarray}\label{E:inuint}
I(\nu_{\rm obs},z=0)=\frac{1}{4\pi}\int_0^\infty dz'\,\frac{dl}{dz'}p[\nu_{\rm obs}(1+z'),z']\, ,
\end{eqnarray}
where $dl/dz=c/[(1+z)H(z)]$ is the proper (not comoving) line element and $p(\nu,z)$ is the emissivity of luminous sources.  For Ly$\alpha$ line emission, the $\nu$-dependence of the emissivity can be written as a delta function in the form
\begin{eqnarray}
p(\nu,z) = \rho_L(z)\delta(\nu-\nu_{\rm Ly\alpha})\, ,
\end{eqnarray}
where $\rho_L(z)$ is the luminosity density of Ly$\alpha$ emitters.  After a little algebra, we can write the intensity as
\begin{eqnarray}\label{E:intj}
I(\nu_{\rm obs})=\frac{c}{4\pi}\frac{\rho_L(z_{\rm Ly\alpha})}{\nu_{\rm Ly\alpha}H(z_{\rm Ly\alpha})}\, ,
\end{eqnarray}
where $z_{\rm Ly\alpha}=\nu_{\rm Ly\alpha}/\nu_{\rm obs}-1$.

The luminosity density $\rho_L(z)$ can be determined directly from the measured LF at redshift $z$.  In particular, the luminosity function $\Phi(L)$, being the number density of emitters per luminosity bin, can produce a luminosity density by integrating $\Phi(L)$ over $L$.  If $\Phi(L)$ is parameterized in the Schechter form \citep{1976ApJ...203..297S}
\begin{eqnarray}
\Phi(L)dL=\Phi_*\left(\frac{L}{L_*}\right)^\alpha e^{-L/L_*}\frac{dL}{L_*}\, ,
\end{eqnarray}
with $\Phi_*$, $L_*$, and $\alpha$ being $z$-dependent, we can write $\rho_L$ as
\begin{eqnarray}
\rho_L=\Phi_*L_*\Gamma[\alpha+2,L_{\rm lim}/L_*]\, ,
\end{eqnarray}
where $\Gamma(\alpha,x)$ is an incomplete gamma function and $L_{\rm lim}$ is the minimum luminosity of the emitters.

We determine $\rho_L(z)$ using the set of LF measurements tabulated by \citet{2011ApJ...730....8H}.  Instead of using their tabulated values of  $\rho_L$, which assume nonzero values of $L_{\rm lim}$, we set $L_{\rm lim}=0$.  The advantage is that it includes unresolved sources, which is important for intensity mapping.  The disadvantage is that it probably overestimates $\Phi(L)$ at small luminosities, making it slightly optimistic.  Note that the various LFs compiled by \citet{2011ApJ...730....8H} have various methods for incompleteness corrections and data reduction, which limits the derived intensity's accuracy over redshift.  We list the LF parameters and $\rho_L$ values in Table~\ref{T:lfpar}.
\begin{table*}[t]
\caption{\label{T:lfpar} Ly$\alpha$ luminosity function parameters from various measurements up to $z\sim8$ with 1-$\sigma$ errors, as well as their inferred luminosity density ($\rho_L$).  Note that the ``errors'' for $\rho_L$ are not truly 1-$\sigma$ errors; they are maximum positive and negative deviations from the expected value based on the 1-$\sigma$ errors of the LF parameters.  Also, we do not give errors for $\alpha$ since this parameter was fixed for most of the LF estimates.}
\begin{center}
\item[]\begin{tabular}{@{}cccccc}
\toprule\toprule Source&$z$&$\alpha$&$L_*$ (10$^{42}$ erg s$^{-1}$)&$\Phi_*$ (10$^{-4}$ Mpc$^{-3}$)&$\rho_L$ (10$^{39}$ erg s$^{-1}$ Mpc$^{-3}$)\\
\midrule
\citet{2008ApJ...680.1072D}&0.275&-1.35&0.955$_{-0.179}^{+0.220}$&3.98$_{-1.23}^{+1.77}$&0.526$_{-0.230}^{+0.409}$\\
\citet{2010ApJ...711..928C}&0.3&-1.36&0.575$_{-0.097}^{+0.116}$&1.95$\pm0.35$&0.157$_{-0.050}^{+0.054}$\\
\citet{Cassata:2011}&2.5&-1.6&5.01&7.1$_{-1.8}^{+2.4}$&7.89$_{-2.00}^{+2.67}$\\
\citet{2007ApJ...667...79G}&3.1&-1.49&4.37$_{-1.27}^{+3.58}$&15.2$\pm1.3$&11.5$_{-4.1}^{+11.3}$\\
\citet{2008ApJS..176..301O}&3.1&-1.5&5.8$_{-0.7}^{+0.9}$&9.2$_{-2.1}^{+2.5}$&9.5$_{-3.0}^{+4.4}$\\
\citet{2008ApJS..176..301O}&3.7&-1.5&10.2$_{-1.5}^{+1.8}$&3.4$_{-0.9}^{+1.0}$&6.1$_{-2.3}^{+3.2}$\\
\citet{Cassata:2011}&3.8&-1.78&5.01&4.8$\pm0.8$&10.$\pm1.7$\\
\citet{2007ApJ...671.1227D}&4.5&-1.6&10.9$\pm3.3$&1.7$\pm0.2$&4.1$_{-1.6}^{+1.9}$\\
\citet{2009ApJ...696..546S}&4.86&-1.5&7.94$_{-3.96}^{+17.18}$&1.2$_{-1.1}^{+8.0}$&1.7$_{-1.6}^{+39.3}$\\
\citet{Cassata:2011}&5.65&-1.69&5.25$\pm1.3$&9.2$_{-1.9}^{+2.3}$&14.$_{-6.}^{+8.}$\\
\citet{2008ApJS..176..301O}&5.7&-1.5&6.8$_{-2.1}^{+3.0}$&7.7$_{-3.9}^{+7.4}$&9.3$_{-6.1}^{+16.9}$\\
\citet{2010ApJ...723..869O}&6.6&-1.5&4.4$\pm0.6$&8.5$_{-2.2}^{+3.0}$&6.6$_{-2.4}^{+3.6}$\\
\citet{Hibon:2010}&7.7&-1.5&10.0$_{-5.0}^{+5.8}$&1.3$_{-0.6}^{+2.7}$&2.3$_{-1.7}^{+8.9}$\\
\bottomrule
\end{tabular}
\end{center}
\end{table*}

\section{Minimum Lyman-alpha Emission - Theoretical} \label{S:modint}

We now attempt to model the intensity of the Ly$\alpha$ line on the sky for both the halo and the IGM based on first principles.  Ly$\alpha$ emission is sourced by UV photons from stars that ionize the surrounding neutral hydrogen.  The resulting free electrons and hydrogen ions then interact with each other and neutral hydrogen to produce hydrogen lines.  While most lines are then reabsorbed by the IGM and lost, Ly$\alpha$ emission is different in that it is both absorbed and re-emitted, causing no net loss in photons.  Thus, we expect to see Ly$\alpha$ emission at very high redshifts.

\begin{figure}
\begin{center}
{\scalebox{0.6}{\includegraphics{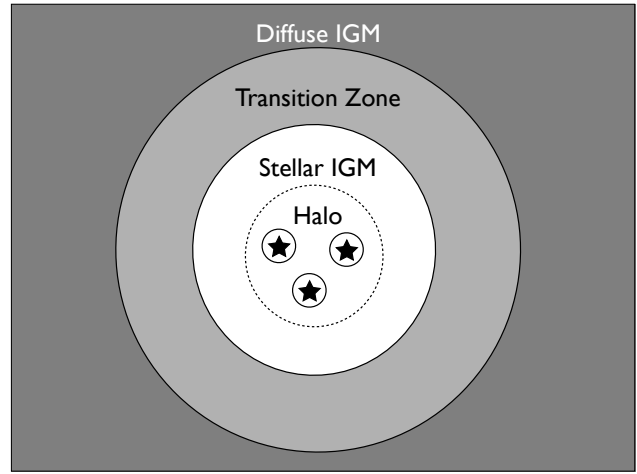}}}
\caption{\label{F:halofig} A cartoon of the emission region around a massive halo, surrounded by the IGM.  The four regions are (starting from the inside) the halo, the stellar IGM, the transition zone, and the diffuse IGM.  The first two regions are ionized, the transition zone is partially ionized, and the diffuse IGM is neutral.  The halo region is denoted by a dashed circle, though the actual regions in ionization equilibrium are denoted as solid circles around the stars.  This figure does not include any ions in the diffuse IGM because it is not part of our fiducial model, even though we consider it in our analysis.}
\end{center}
\end{figure}

We model the Ly$\alpha$ emitter as a galaxy with a Pop II or Pop III stellar population within a massive halo surrounded by the IGM.  The Ly$\alpha$ emitter can be divided into three parts: the massive halo, the stellar IGM, and the transition zone, as shown in Fig.~\ref{F:halofig}.  The massive halo contains the stellar population, is much more dense than the mean cosmological density, and is 100\% ionized, with each star surrounded by a volume exhibiting ionization equilibrium as a Str\"{o}mgren sphere \citep{2002MNRAS.336.1082S}.  The stellar IGM surrounds the halo, has the mean cosmological density, and is also ionized, though not in ionization equilibrium.  Surrounding this is the transition zone, in which the ionization fraction transitions from fully ionized to fully neutral, with a region size determined by the mean free path of ionizing photons.  The volumes of the halo and stellar IGM ionized regions, denoted by $V_{\rm halo}$ and $V_{\rm IGM}$, and the transition zone's size $S_{\rm TZ}$ are given by
\begin{eqnarray}
V_{\rm halo}(m) = \frac{(1-f_{\rm esc})\overline{Q}_H(m)}{n_en_p\alpha_B}\, ,
\end{eqnarray}
\begin{eqnarray}
V_{\rm IGM}(m) = \frac{f_{\rm esc}\overline{Q}_H(m)\tau(m)}{n_H}\, ,
\end{eqnarray}
and
\begin{eqnarray} \label{E:stz}
S_{\rm TZ} = \frac{1}{n_H\sigma_{\VEV{\nu}}}\, ,
\end{eqnarray}
where $n_e=n_p=10^4$ cm$^{-3}$ are the number densities of electrons and protons in the halo, $n_{H}(z)=1.905\times10^{-7}$ cm$^{-3}$ is the proper number density of all hydrogen, $\alpha_B$ is the case-B recombination coefficient, and $\sigma_{\VEV{\nu}}$ is the photoionization cross section.  The parameter $f_{\rm esc}$ is the fraction of ionizing photons that escape from the halo to emit from the IGM, and $\overline{Q}_H(m)$ is the production rate of ionizing photons, and $\tau(m)$ is the stellar lifetime, given by the fitting formulas in \citet{2002A&A...382...28S},
\begin{eqnarray}
&&\log_{10}\left(\frac{\overline{Q}_H}{\rm s^{-1}}\right) =
             \left\{\begin{array}{ll}
             39.29+8.55x, & m=5-9M_\odot\\
             43.61+4.90x-0.83x^2, &\\
  \,\,\,\,\,\,\,\,\,\,\,\,\,\,\,\,\,\,\,\,\,\,\,\,\,\,\,\,\,\,\,\,\,\,\,\,\,\,\,\,\,\,\,\,\,\,m=9&-500M_\odot\end{array}\right.\nonumber\\
&&\log_{10}\left(\frac{\tau}{\rm yr}\right) = 9.785-3.759x+1.413x^2-0.186x^3\, ,
\end{eqnarray}
with $5M_\odot<m<500M_\odot$ for Pop III stars and
\begin{eqnarray}
\log_{10}\left(\frac{\overline{Q}_H}{\rm s^{-1}}\right) &=& 27.80+30.68x-14.80x^2+2.50x^3\nonumber\\
\log_{10}\left(\frac{\tau}{\rm yr}\right) &=& 9.59-2.79x+0.63x^2\, ,
\end{eqnarray}
with $7M_\odot<m<150M_\odot$ for Pop II stars, where $x\equiv\log_{10}(m/M_\odot)$.  We set $Q(m)$ to zero outside the given mass ranges.

The mean intensity $I(\nu_{\rm obs})$ can be given in terms of the {\it comoving} volume emissivity $p(\nu,z)$, as in Eq.~\ref{E:inuint}, according to
\begin{eqnarray}\label{E:intcalc}
I(\nu_{\rm obs}) = \frac{c}{4\pi}\int dz\,\frac{p[\nu_{\rm obs}(1+z),z]}{H(z)(1+z)}\, .
\end{eqnarray}
The {\it stellar} comoving volume emissivity for both the halo and the IGM is given by
\begin{eqnarray}\label{E:emis}
p(\nu,z) = \frac{\dot{\rho_*}(z)}{m_*}\int dm\,f(m)\overline{L}_\nu(m)\tau(m)\, ,
\end{eqnarray}
where $m_*$ is the mean stellar mass given by the first moment of $f(m)$, the initial mass function (IMF), and $\tau(m)$, the stellar main-sequence lifetime.  Note that this expression is multiplied by $(1-f_{\rm dust})$ for the halo, which represents the fraction of Ly$\alpha$ halo emission \emph{not} obscured by dust.  

\begin{figure}
\begin{center}
{\scalebox{0.5}{\includegraphics{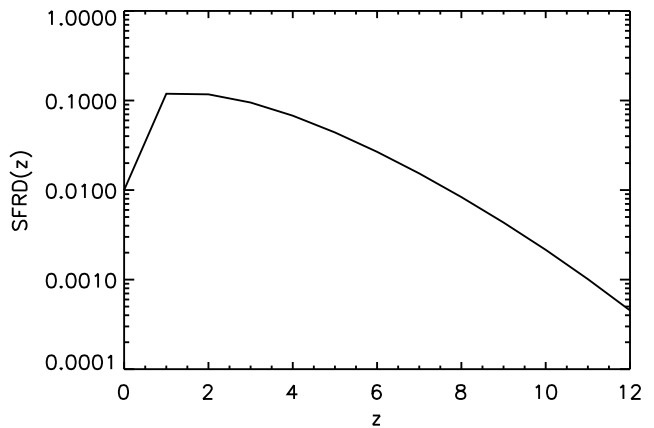}}}
\caption{\label{F:sfrd} The star formation rate density used in our analysis.}
\end{center}
\end{figure}

The star formation rate (SFR) can be parametrized in the simple form \citep{2005ApJ...620..553L}
\begin{eqnarray}
SFR=f_*(M)\frac{\Omega_b}{\Omega_m}\frac{M}{t_s}\, ,
\end{eqnarray}
where $f_*(M)$, the star formation efficiency, is the fraction of baryons that comprise stars, $M$ is the halo mass, and $t_s$ is the star formation timescale.  From \citet{2003ApJ...586..693W} Eq.~13, we use the model $f_*(M)\propto M^{2/3}$ below $M_*$ and constant above it, suggested by $z\sim 0$ observations of \citet{2003MNRAS.341...54K}.  Using this SFR model, we write the SFRD, which is just the integral of the SFR over the halo mass function times $f_{\rm duty}=t_s/t_{\rm age}(z)$, as
\begin{eqnarray}
\dot{\rho_*}(z) &=& 0.234\,M_\odot{\rm yr^{-1}Mpc^{-3}}\nonumber\\
&&\times\left[\frac{t_{\rm age}(z=2)}{t_{\rm age}(z)}\right]\left[\frac{q_{\rm coll}(M_{\rm min},z)}{0.126}\right]\, ,
\end{eqnarray}
where $q_{\rm coll}(M_{\rm min},z)$ is the fraction of {\it star formation} in collapsed halos with masses greater than $M_{\rm min}$, given by
\begin{eqnarray}
q_{\rm coll}=\frac{1}{\rho_m}\int_{M_{\rm min}}^{M_{\rm max}} dM\,n(M)Mf_*(M)\, ,
\end{eqnarray}
where we assume the Tinker fitting formula \citep{2008ApJ...688..709T} for the halo mass function $n(M)$, $t_{\rm age}$ is the age of the universe at redshift $z$, $\rho_m$ is the mean matter density today,  and $M_{\rm min}$ and $M_{\rm max}$ are the minimum and maximum halo masses, respectively, that we consider in our model.  We set the minimum halo mass to $M_{\rm min}=10^9M_\odot$ because less massive halos cannot produce stars efficiently \citep{2010ApJ...710.1089F}; we also set the maximum halo mass equal to $M_{\rm max}=10^{15}M_\odot$.  We also set $f_*(M\geq M_*)=0.5$ throughout the rest of the modeling to match the SFRD measurement at $z=2$ in \citet{2006ApJ...651..142H}.  While this model reproduces measured values fairly well over redshifts $z>1$, the values for $z<1$ appeared to be too high.  Thus, we use the value $\dot{\rho}_*(z=0)\simeq0.01\,{\rm M_\odot\,yr^{-1}\,Mpc^{-3}}$ from \citet{2006ApJ...651..142H}.  We plot the SFRD we use in Fig.~\ref{F:sfrd}.  Similar to LFs, the SFR is also calibrated to bright objects; thus, the SFR we use is a lower limit of the true SFR including lower mass objects, meaning the resulting Ly$\alpha$ emission can indeed be higher than our estimates.

Although $f_{\rm esc}$ is redshift- and halo-mass-dependent, the parameter is constrained to $f_{\rm esc}<0.1$ for low redshifts ($z\lesssim 6$).  Thus, we choose to set $f_{\rm esc}=0.1$ for redshifts $z\leq6$ since the halo emission is barely affected and the stellar IGM emission, which is subdominant to the halo emission at these redshifts, is set to an upper limit.  For higher redshifts, $f_{\rm esc}$ does deviate significantly from 0.1.  Its values are very uncertain, which will reflect in the uncertainty of our theoretical model.  To construct representative values of $f_{\rm esc}$, we take the values for $f_{\rm esc}(M,z)$, where M is the halo mass, from simulations constructed in \citet{2010ApJ...710.1239R}, and average them over the halo mass function $n(M)$ times the SFR factors $f_*(M)M$, according to
\begin{eqnarray}
f_{\rm esc}(z) = \frac{\int_{M_{\rm min}}^{M_{\rm max}}dM\,n(M)Mf_*(M)f_{\rm esc}(M,z)}{\int_{M_{\rm min}}^{M_{\rm max}}dM\,n(M)Mf_*(M)}\, .
\end{eqnarray}
The values for redshifts $z\geq6$ are listed in Table \ref{T:fesc}.
\begin{table}
\caption{\label{T:fesc} Values of $f_{\rm esc}$ used for redshifts $z\geq6$.}
\begin{center}
\item[]\begin{tabular}{@{}cc}
\toprule\toprule $z$&$f_{\rm esc}$\\
\midrule
6&0.100\\
7&0.171\\
8&0.387\\
9&0.649\\
10&0.735\\
11&0.757\\
12&0.763\\
\bottomrule
\end{tabular}
\end{center}
\end{table}

\citet{2011ApJ...730....8H} estimated the dust obscuration by comparing observed Ly$\alpha$ and H$\alpha$ emission, deriving the fitting formula
\begin{eqnarray}
1-f_{\rm dust}=C_{\rm dust}\times10^{-3}(1+z)^\xi\, ,
\end{eqnarray}
with $C_{\rm dust}=1.67$ and $\xi=2.57$.  The luminosity parameter $\overline{L}_\nu(m)$ will be explained later in this section.

The stellar distribution in the low redshift halos consists of mainly two components, the disk and the stellar halo.  We will use Chabrier stellar halo IMF \citep{2003PASP..115..763C} for Pop II stars, given in the log-normal form
\begin{eqnarray}
f(m)\propto \left\{\begin{array}{ll}
             m^{-1}\exp\left[-\frac{(\log m-\log m_c)^2}{2\sigma^2}\right] & m\leq m_f\\
             Am^{-2.3} & m > m_f\end{array}\right.\, ,
\end{eqnarray}
with $m$ in units of $M_\odot$ and the parameters given by $m_f=0.7M_\odot$, $m_c=0.22M_\odot$, $\sigma=0.33$, and $A=0.20$.  At high redshifts, Pop III exhibit a different stellar IMF, parametrized as
\begin{eqnarray}
f(m)\propto m^{-1.9}\exp\left[-\left(\frac{3.2M_\odot}{m}\right)^{1.6}\right]\, ,
\end{eqnarray}
in \citet{2003PASP..115..763C}.

The individual luminosity functions of the halos and the IGM represent how ionizing photons from Pop III stars radiate to produce halo and IGM emission.  The luminosity $\overline{L}_\nu(m)$ is given by the product of the Ly$\alpha$ proper emissivity of ionized particles and the proper volume of the Str\"{o}mgren sphere that gets ionized (assuming equilibrium).  The proper emissivity $p_{\rm Ly\alpha}$ is given by
\begin{eqnarray}
p_{\rm Ly\alpha}(\nu,z) &=& h_P\nu_{\rm Ly\alpha}[n_en_p f_{\rm Ly\alpha}(T_{\rm gas})\alpha_B(T_{\rm gas})\nonumber\\
&&+n_en_{HI}q_{\rm eff}(T_{\rm gas})]\phi(\nu_{\rm Ly\alpha}-\nu)\, ,
\end{eqnarray}
where $h_P$ is Planck's constant, $f_{\rm Ly\alpha}$ is the fraction of $n=2\to1$ emissions that are Ly$\alpha$ photons, $\alpha_B$ is the case-B recombination coefficient, $q_{\rm eff}$ is the effective collisional excitation coefficient, and $n_e$, $n_p$, and $n_{HI}$ are proper number densities of electrons, protons, and neutral hydrogen, respectively.  The first term in the bracket is due to recombinations and the second term is due to collisions.  The expressions for $f_{\rm Ly\alpha}(T_{\rm gas})$, $\alpha_B(T_{\rm gas})$, and $q_{\rm eff}(T_{\rm gas})$ can be found in \citet{2008ApJ...672...48C}.  $\phi(\nu_{\rm Ly\alpha}-\nu,z)$ is the line profile.  Although the line profile is finite due to absorption in the halo,  we will neglect the line width in the analysis and set $\phi(\nu_{\rm Ly\alpha}-\nu,z)=\delta_D(\nu-\nu_{\rm Ly\alpha})$.

\subsection{Halo contribution}\label{S:halo}
The halo luminosity for the Ly$\alpha$ line due to recombinations is given by
\begin{eqnarray}
\overline{L}_\nu^{\rm halo}(m,z) = h_P\nu_{\rm Ly\alpha}\overline{Q}_H(m)f_{\rm Ly\alpha}\delta_D(\nu_{\rm Ly\alpha}-\nu)\, .
\end{eqnarray}
The emissivity can then be written in the form
\begin{eqnarray}
p_{\rm halo}(\nu,z) &=& (1-f_{\rm dust})(1-f_{\rm esc})f_{\rm Ly\alpha}h_P\nu_{\rm Ly\alpha}\delta_D(\nu_{\rm Ly\alpha}-\nu)\nonumber\\
&&\times\frac{\dot{\rho_*}(z)}{m_*}\int dm\,f(m)\overline{Q}_H(m)\tau(m)\, .
\end{eqnarray}
When inserting this into the mean intensity integral, we take advantage of the fact that the line profile is a delta function.  This makes the integral simple, giving the form of the mean intensity as
\begin{eqnarray}
I_{\rm halo}^{\rm Ly\alpha}(z_{\rm Ly\alpha})&=&[1-f_{\rm dust}(z_{\rm Ly\alpha})][1-f_{\rm esc}(z_{\rm Ly\alpha})]f_{\rm Ly\alpha}\nonumber\\
&&\times\frac{ch_P}{4\pi H(z_{\rm Ly\alpha})}\dot{\rho_*}(z_{\rm Ly\alpha})Q_{\rm ion}^1(z_{\rm Ly\alpha})\, ,
\end{eqnarray}
where $Q_{\rm ion}^1$ is the stellar mass integral divided by $m_*$ and $z_{\rm Ly\alpha}=\nu_{\rm Ly\alpha}/\nu_{\rm obs}-1$ is the initial redshift of the Ly$\alpha$ photon.

The only temperature-dependent quantity, $f_{\rm Ly\alpha}$, varies very little with temperature.  Thus, we will assume the gas temperature in the halo $T_{\rm gas}^{\rm halo}=2\times10^4$K, an expected gas temperature in a massive halo that can become dense enough through atomic line cooling to form stars \citep{2007ApJ...671....1T}.  For this temperature, we can set $f_{\rm Ly\alpha}=0.64$.  $Q_{\rm ion}^1$ is of order $10^{60}\,M_\odot^{-1}$ for the models we consider.  We give the values for the various stellar population models in Table \ref{T:qion}.  In our model, similarly to \citet{2012ApJ...756...92C}, we assume that Pop II stars dominate the signal for $z<10$ and that Pop III stars dominate for $z>10$, which we implement by constructing the toy model $Q_{\rm ion}^1(z) = Q_{\rm ion,Pop\,II}^1(1-f_P(z))+Q_{\rm ion,Pop\,III}^1f_P(z)$, where
\begin{eqnarray}
f_P(z) = \frac{1}{2}\left[1+\tanh\left(\frac{z-10}{0.5}\right)\right]\, .
\end{eqnarray}
Substituting these values gives the expression for our case
\begin{eqnarray}\label{E:Lyalint2}
I_{\rm halo}^{\rm Ly\alpha}(z_{\rm Ly\alpha})&=&0.0049\,{\rm nW\,m^{-2}\,THz^{-1}\,str^{-1}}(1-f_{\rm dust})\nonumber\\
&&\times(1-f_{\rm esc})\left(\frac{H(z_{\rm Ly\alpha}=6)}{H(z_{\rm Ly\alpha})}\right)\left(\frac{Q_{\rm ion}^1}{10^{60}\,M_\odot^{-1}}\right)\nonumber\\
&&\times\left(\frac{\dot{\rho_*}(z_{\rm Ly\alpha})}{10M_\odot{\rm yr^{-1}Mpc^{-3}}}\right)\, .
\end{eqnarray}

An equivalent way to write this result is to express the luminosity density of Ly$\alpha$ photons as
\begin{eqnarray}
\rho_L(z)=f_{\rm Ly\alpha}(1-f_{\rm dust})(1-f_{\rm esc})h_P\nu_{\rm Ly\alpha}\dot{n}_{\rm ion}(z)\, ,
\end{eqnarray}
where $n_{\rm ion}(z)$ is the rate per comoving volume of ionizing photons produced by stars in the halo, given by
\begin{eqnarray}
\dot{n}_{\rm ion}(z)=Q_{\rm ion}^1(z)\dot{\rho_*}(z)\, ,
\end{eqnarray}
which can be converted to an intensity using Eq.~\ref{E:intj}.  It is possible that the SFRD used in our analysis does not include all stars that contribute Ly$\alpha$ photons.  In particular, we may be missing the contribution from low-mass stars and dwarfs.  We can set a lower limit to $\dot{n}_{\rm ion}$ by requiring the rate per comoving volume of photons that ionize the IGM $f_{\rm esc}\dot{n}_{\rm ion}$ to be greater than $\dot{n}_{\rm rec}$, the rate per comoving volume of recombinations in the IGM, in order for net reionization to be sustained.  The recombination rate can written as \citep{1999ApJ...514..648M,2011MNRAS.414..847S}
\begin{eqnarray}
\dot{n}_{\rm rec}=10^{50.0}C(z)\left(\frac{1+z}{7}\right)^3\,{\rm s^{-1}Mpc^{-3}}\, ,
\end{eqnarray}
where $C(z)=\VEV{n_e^2}/\VEV{n_e}^2$ is the clumping factor for ionized hydrogen, which we set to $C(z)=26.2917 e^{-0.1822z+0.003505z^2}$ \citep{2007MNRAS.376..534I}.  By setting $\dot{n}_{\rm rec}/f_{\rm esc}$ as a lower limit to $\dot{n}_{\rm ion}$, we find that the intensity of Ly$\alpha$ photons cannot be less than
\begin{eqnarray}
I_{\rm halo,\,min}^{\rm Ly\alpha}(z_{\rm Ly\alpha})&=&1.55\times10^{-5}\,{\rm nW\,m^{-2}\,THz^{-1}\,str^{-1}}\nonumber\\
&&\times\left(\frac{1-f_{\rm esc}}{f_{\rm esc}}\right)\left(\frac{H(6)}{H(z_{\rm Ly\alpha})}\right)\left(\frac{C(z_{\rm Ly\alpha})}{C(6)}\right)\nonumber\\
&&\times\left(\frac{1+z_{\rm Ly\alpha}}{7}\right)^3(1-f_{\rm dust})\, 
\end{eqnarray}

\begin{figure}
\begin{center}
{\scalebox{0.5}{\includegraphics{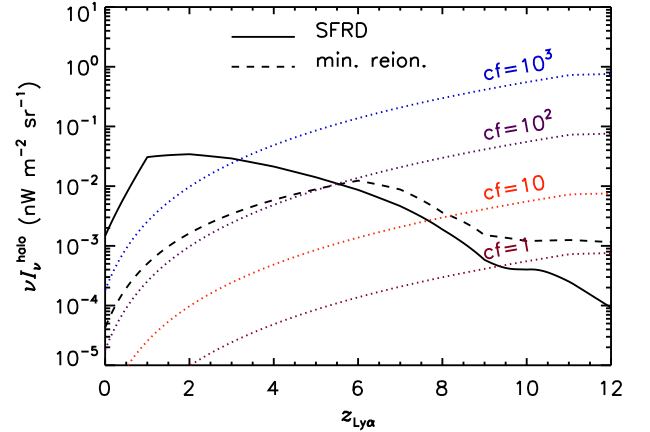}}}
\caption{\label{F:halo} The Ly$\alpha$ intensity due to halo emission, based on our SFRD model (solid) and the minimum ionizing photon production needed to sustain reionization (dashed), as well as for various values of $cf=C(1-f_{\rm esc})/f_{\rm esc}$.  We choose the SFRD estimate for $z_{\rm Ly\alpha}<5.5$ and minimum reionization estimate for higher redshifts.}
\end{center}
\end{figure}

We plot both estimates of the Ly$\alpha$ emission from halos in Fig.~\ref{F:halo}, and we see that for redshifts $z_{\rm Ly\alpha}\gtrsim5.5$, the minimum required emission is greater than the estimated value based on our SFRD.  This can greatly impact the total emission in our model, particularly at very high redshifts.  Thus, in our analysis we will use the SFRD model for lower redshifts where it is greater, but we will use the minimum reionization value at higher redshifts.  Note that this is indeed a lower limit; the actual emission can be even higher, particularly closer to the completion of reionization where we would expect the IGM to receive an excess of ionizing photons.  Also, these minimum levels are very dependent on the values of the clumping factor and the escape fraction.  To illustrate this, we also plot in Fig.~\ref{F:halo} the minimum intensity predicted for various values of $cf=C(1-f_{\rm esc})/f_{\rm esc}$, where we see that this minimum reionization condition kicks at earlier redshifts for higher values of $cf$.

\citet{2013ApJ...763..132S} point out that there are other sources of Ly$\alpha$ photons in halos that can contribute to the intensity mapping signal.  These include excitations in the hydrogen gas, gas cooling, and cascading Ly-$n$ photons from continuum emission (within the halo).  We decline to include these mechanisms because they are all negligible compared to the signal from recombinations.  The highest of these other contributions, namely excitations, is about an order of magnitude below the recombinations signal because only 10\% of the excitation energy due to collisions, which is close to the Ly$\alpha$ energy, is emitted as Ly$\alpha$ photons, whereas halo emission emits almost all its ionizing energy.
\begin{table}[!t]
\begin{center}
\caption{\label{T:qion} $Q_{\rm ion}^1$, $Q_{\rm ion}^2$, and $T_{\rm ion}$ for various stellar metallicities.}
\begin{tabular}{cccc}
\hline
Metallicity&$Q_{\rm ion}^1$&$Q_{\rm ion}^2$&$T_{\rm ion}$\\
& $[10^{60}M_\odot^{-1}]$ & $[10^{67}M_\odot^{-1}{\rm yr}]$ & $[10^5M_\odot^{-1}{\rm yr}]$ \\
\hline
Pop II&8.3&4.5&2.3\\
Pop III&23.&18.&16.\\
\hline
\end{tabular}\end{center}
\end{table}

\subsection{IGM contribution from ionizing stars}\label{S:igms}
Ionizing photons from stars that escape their halos can ionize neutral hydrogen in the IGM.  The resulting ions and electrons can produce Ly$\alpha$ emission through recombination, similar to halo emission. The IGM luminosity from stars can be given as
\begin{eqnarray}
\overline{L}_\nu^{\rm igm*}(m,z) &=& f_{\rm esc}h_P\nu_{\rm Ly\alpha}\frac{\overline{Q}_H(m)\tau(m)}{n_H}n_en_pf_{\rm Ly\alpha}\alpha_B\nonumber\\
&&\times\delta_D(\nu_{\rm Ly\alpha}-\nu)\, ,
\end{eqnarray}
where we assume $n_e=n_p=n_H$ due to charge neutrality.

This parametrization gives the emissivity in the form
\begin{eqnarray}
p_{\rm igm*}(\nu,z) &=& f_{\rm esc}f_{\rm Ly\alpha}\alpha_Bh_P\nu_{\rm Ly\alpha}\phi(\nu_{\rm Ly\alpha}-\nu,z)C(z)n_H(z)\nonumber\\
&&\times\frac{\dot{\rho_*}(z)}{m_*}\int dm\,f(m)\overline{Q}_H(m)\tau^2(m)\, .
\end{eqnarray}

When calculating the line intensity, we treat the line profile in the same manner as the previous section; therefore, our expression for the line intensity due to the IGM can be written in the form
\begin{eqnarray}
I_{\rm igm*}^{\rm Ly\alpha}(z_{\rm Ly\alpha}) &=&  f_{\rm esc}f_{\rm Ly\alpha}\alpha_B\frac{h_Pc}{4\pi H(z_{\rm Ly\alpha})}C(z_{\rm Ly\alpha})n_H(z_{\rm Ly\alpha})\nonumber\\
&&\times\dot{\rho_*}(z_{\rm Ly\alpha})Q_{\rm ion}^2\, ,
\end{eqnarray}
where $Q_{\rm ion}^2$ is the stellar mass integral in this equation divided by $m_*$, which is on the order of $10^{67}\,M_\odot^{-1}$ yr for the models we consider.  We vary $Q_{\rm ion}^2$ with redshift in a similar manner to $Q_{\rm ion}^1$ in the previous section.

We assume an IGM gas temperature of $T_{\rm gas}^{\rm halo}=2\times10^4$K, which sets $f_{\rm Ly\alpha}\alpha_B=9.25\times10^{-14}\,{\rm cm^3\,s^{-1}}$.  We will ignore the perturbations in $n_H$ (which may underestimate the signal) and set $n_H(z)=1.905\times10^{-7}(1+z)^3\,{\rm cm^{-3}}$, the mean hydrogen number density. Substituting these values gives us
\begin{eqnarray}
I_{\rm igm*}^{\rm Ly\alpha}(z_{\rm Ly\alpha}) &=&1.45\times10^{-3}\,{\rm nW\,m^{-2}\,THz^{-1}\,str^{-1}}f_{\rm esc}\nonumber\\
&&\times\left(\frac{1+z_{\rm Ly\alpha}}{7}\right)^3\left(\frac{f_{\rm Ly\alpha}\alpha_B}{9.25\times10^{-14}{\rm cm^3s^{-1}}}\right)\nonumber\\
&&\times\left(\frac{C(z_{\rm Ly\alpha})}{C(6)}\right)\left(\frac{Q_{\rm ion}^2}{10^{68}\,M_\odot^{-1}{\rm yr}}\right)\nonumber\\
&&\times\left(\frac{\dot{\rho_*}(z_{\rm Ly\alpha})}{10{\rm M_\odot yr^{-1}Mpc^{-3}}}\right)\left(\frac{H(6)}{H(z_{\rm Ly\alpha})}\right)\, .
\end{eqnarray}

\subsection{Transition zone emission}\label{S:trans}

Outside the stellar IGM, the ionization state transitions from fully ionized to partially ionized.  In this region, electrons, protons, and neutral hydrogen exists, producing emission from recombinations and collisional excitations.  At our fiducial temperature of $T_{\rm gas}=2\times10^4$ K, collisional excitations will dominate the spectrum.  The luminosity of this region is given by
\begin{eqnarray}
\overline{L}_\nu^{\rm tz}(m,z) &=& h_P\nu_{\rm Ly\alpha}q_{\rm eff}\delta_D(\nu_{\rm Ly\alpha}-\nu)\nonumber\\
&&\times4\pi\int_0^{S_{\rm TZ}} dr\,r^2n_e(r)n_{HI}(r)\nonumber\\
&=& h_P\nu_{\rm Ly\alpha}q_{\rm eff}C(z)n_H^2(z)\delta_D(\nu_{\rm Ly\alpha}-\nu)\nonumber\\
&&\times4\pi\int_0^{S_{\rm TZ}} dr\,r^2x_e(r)[1-x_e(r)]\, ,
\end{eqnarray}
where $S_{\rm TZ}$ is given in Eq.~\ref{E:stz}.  We approximate the ionization fraction as a linear function with $x_e(r=0)=1$ and $x_e(r=S)=0$, getting an integral value of $S_{\rm TZ}^3/20$.  Note that it would be more precise to integrate from the end of the ionized region, not $r=0$, but the emission volume, which just depends on the number of atoms that get ionized, should not depend on the inner radius of the transition zone.  Thus, our approximation should not affect the integration much, although the distribution of $x_e$ in the volume will be dependent on this radius.

Similar to the previous sections, we can convert this to an emissivity and then an intensity given by
\begin{eqnarray}
I_{\rm tz}^{\rm Ly\alpha}(z_{\rm Ly\alpha}) &=& \frac{h_Pc}{H(z_{\rm Ly\alpha})}C(z_{\rm Ly\alpha})n_H^2(z_{\rm Ly\alpha})q_{\rm eff}\nonumber\\
&&\times \frac{S_{\rm TZ}^3}{20}\dot{\rho_*}(z_{\rm Ly\alpha})T_{\rm ion}\, ,
\end{eqnarray}
where
\begin{eqnarray}
T_{\rm ion} = \frac{1}{m_*}\int dm\,f(m)\tau(m)\, .
\end{eqnarray}
Our fiducial temperature sets $q_{\rm eff} = 5.27\times10^{-11}\,{\rm cm^3 s^{-1}}$.  The mean free path, assuming $\VEV{\nu} = 3\nu_{LL}$, the Lyman-limit frequency, is $1.46(1+z)^{-3}$ Mpc.  We also fiducially set $T_{\rm ion} = 10^5\,{\rm yr M_\odot^{-1}}$.  This gives us
\begin{eqnarray}
I_{\rm tz}^{\rm Ly\alpha}(z_{\rm Ly\alpha}) &=&7.813\times10^{-10}\,{\rm nW\,m^{-2}\,THz^{-1}\,str^{-1}}\nonumber\\
&&\times\left(\frac{7}{1+z_{\rm Ly\alpha}}\right)^3\left(\frac{q_{\rm eff}(T_{\rm gas})}{5.27\times10^{-11}\,{\rm cm^3 s^{-1}}}\right)\nonumber\\
&&\times\left(\frac{C(z_{\rm Ly\alpha})}{C(6)}\right)\left(\frac{S(1+z_{\rm Ly\alpha})^3}{1.46\,{\rm Mpc}}\right)^3\nonumber\\
&&\times\left(\frac{\dot{\rho_*}(z_{\rm Ly\alpha})}{10{\rm M_\odot yr^{-1}Mpc^{-3}}}\right)\left(\frac{H(6)}{H(z_{\rm Ly\alpha})}\right)\nonumber\\
&&\times\left(\frac{T_{\rm ion}}{10^5\,{\rm yr M_\odot^{-1}}}\right)\, .
\end{eqnarray}
This emission is only nonnegligible at very low redshifts ($z\lesssim1$).  However, at these redshifts all the IGM should be ionized, eliminating any transition zones.  This formalism doesn't account for this effect.  Thus, we will neglect this emission for the rest of this analysis.

\subsection{Diffuse IGM contribution from recombinations}
Separate from the line emission resulting from discrete Ly$\alpha$ emitters are diffuse components to Ly$\alpha$ emission from the IGM.  The first contribution we consider results from recombinations and collisions of electrons and H-ions {\it apart} from ionized halos.  This is a more exotic source of emission, in that inhomogeneous reionization from stellar populations is the fiducial model and new sources of ionization beyond star formation would be required; however, there have been models of decaying or annihilating particles that can produce homogeneous reionization, though these are constrained by CMB experiments \citep{2006PhR...433..181F}.  However, we model the case where an unknown mechanism produces a homogeneous ionization fraction $X_e(z)$ where electrons, electrons, and neutral hydrogen can easily interact.

The diffuse emission can be calculated using Eq.~\ref{E:intcalc}, except that the emissivity is now of the form
\begin{eqnarray}\label{E:igmemis}
p_{\rm diff\,rec/coll}(\nu,z) &=& (1+z)^3h_P\nu_{\rm Ly\alpha}[n_en_p f_{\rm Ly\alpha}(T)\alpha_B(T)\nonumber\\
&&+n_en_{HI}q_{\rm eff}(T)]\phi(\nu_{\rm Ly\alpha}-\nu)\, ,
\end{eqnarray}
where $n_e=n_p=X_en_H$, $n_{HI}=(1-X_e)n_H$, $n_H$ is the \textit{comoving} number density of hydrogen equal to $n_H(z=0)$, and $X_e(z)$ is the ionization fraction.  In our fiducial model, we will set the gas temperature in the IGM $T_{\rm gas}^{\rm IGM}=10^3$ K, though we note that this gives a conservative estimate of the signal in that lower temperatures tend to produce more recombinations.  In this case, $f_{\rm Ly\alpha}=0.77$ and $\alpha_B = 1.49\times10^{-12}\,{\rm cm^3\,s^{-1}}$.  Note that using the case-A recombination coefficient (1.6 times larger than case-B) is more accurate at low redshifts when the IGM is highly dense and ionized \citep{2006PhR...433..181F}.  Thus, we will scale $\alpha$ with $X_e(z)$ as a rough estimate to be case-B at high redshifts and case-A at low redshifts.

\begin{figure}
\begin{center}
{\scalebox{0.5}{\includegraphics{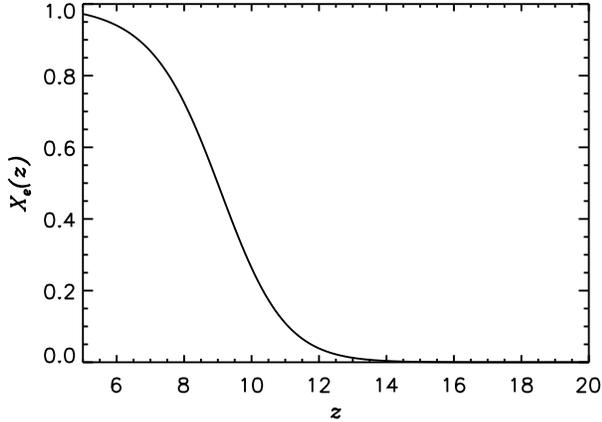}}}
\caption{\label{F:xe} The model for the ionization fraction in the IGM we use in our forecasts.}
\end{center}
\end{figure}

To model the ionization fraction $X_e(z)$, we simply use the toy model assuming a spatially averaged rapid reionization,
\begin{eqnarray}
X_e(z) = \frac{1}{2}\left[1+\tanh\left(\frac{y(z_{\rm re})-y(z)}{\Delta y}\right)\right]\, ,
\end{eqnarray}
where $y(z) = (1+z)^{3/2}$, $\Delta y = 1.5\sqrt{1+z_{\rm re}}\Delta z$, $z_{\rm re}$ is the redshift halfway through the reionization epoch, and $\Delta z$ = $z_{\rm re}-z_{\rm end}$.  We choose reionization to begin at $z=11$ and end at $z=7$ (consistent with WMAP data), giving us $z_{\rm re}=9$ and $\Delta y=9.5$.  We plot $X_e(z)$ in Fig.~\ref{F:xe}.
This is meant only as an illustrative example that matches the data, not a prediction.

Unlike the previous two cases, when we calculated a luminosity function to construct the emissivity, this time we use the emissivity directly since the emission is not associated with stars.  This gives us the expression
\begin{eqnarray}
I_{\rm diff\,rec/coll}^{\rm Ly\alpha}(z_{\rm Ly\alpha}) &=&  \frac{h_Pc(1+z_{\rm Ly\alpha})^3}{4\pi H(z_{\rm Ly\alpha})}X_e^2(z_{\rm Ly\alpha})f_{\rm Ly\alpha}(T)\nonumber\\
&\times& \alpha(T)C(z)n_H^2
\end{eqnarray}
The hydrogen producing this emission is confined to halos that do not produce ionizing stars.  Thus, we write
\begin{eqnarray}
I_{\rm diff\,rec/coll}^{\rm Ly\alpha}(z_{\rm Ly\alpha}) &=& 1.0\times10^{-4}{\rm nW.\,m^{-2}.\,THz^{-1}.\,str^{-1}}\nonumber\\
&&\times\left(\frac{1+z_{\rm Ly\alpha}}{7}\right)^3\left(\frac{H(6)}{H(z_{\rm Ly\alpha})}\right)\left(\frac{C(z)}{C(6)}\right)\nonumber\\
&&\times\left(\frac{X_e^2f_{\rm Ly\alpha}\alpha}{1.15\times10^{-12}{\rm cm^3s^{-1}}}\right)\, .
\end{eqnarray}

\subsection{Diffuse IGM from continuum photons}

The second contribution we consider results from continuum photons from stellar regions that redshift into Lyman-series frequencies, are absorbed by regions of neutral hydrogen, and cascade in the Ly$\alpha$ photons.  This photon emission is important for the Wouthuysen-Field effect \citep{1952AJ.....57R..31W,1959ApJ...129..536F} that couples Ly$\alpha$ photons to 21-cm photons; however, the Ly$\alpha$ photons are an important source of emission in and of themselves.  Several articles give the formula for the intensity \citep{2005ApJ...626....1B,2006MNRAS.367..259H,2008ApJ...689....1S}; however, the formulas assume the high-redshift universe, where there is no ionized hydrogen.  Since we are interested in all cosmic times, we must account for this by including factors denoting the transmission/absorption probabilities.

The relevant formula for our case is given by
\begin{eqnarray}\label{E:Icont}
I_{\rm diff\,cont}^{\rm Ly\alpha}(z) &=& \frac{h\nu_{\rm Ly\alpha}}{4\pi}\sum_{n=2}^{\infty}f_{\rm rec}(n)\int_z^\infty dz'\,\frac{c}{(1+z')H(z')}\nonumber\\
&&\times\dot{\rho_*}(z')\epsilon(\nu_n') P_{\rm abs}(n,z)\nonumber\\
&&\times\prod_{n'=n+1}^{n_{\rm max}}\{1-P_{\rm abs}[n',z_{n'}(z,n)]\}\, .
\end{eqnarray}
The factor $f_{\rm rec}(n)$ is the probability for a Ly$n$ photon to cascade down to a Ly$\alpha$ photon. The continuum spectrum is denoted as $\epsilon(\nu)$, and $\nu_n'$ is the emitted frequency at redshift $z'$ that redshifts to a Ly$n$ photons at redshift $z$, given by
\begin{eqnarray}
\nu_n'(z,z',n) = \nu_{\rm LL}(1-n^{-2})\left(\frac{1+z'}{1+z}\right)\, ,
\end{eqnarray}
where $\nu_{\rm LL}$ is the Lyman-limit frequency and $z_{n'}(z,n)$ gives the redshift of higher Ly$n'$ lines that become Ly$n$, given by
\begin{eqnarray}
1+z_{n'}(z,n) = (1+z)\frac{1-(n')^{-2}}{1-n^{-2}}\, .
\end{eqnarray}
We set the spectrum $\epsilon(\nu)\propto \nu^{-\alpha}$ with $\alpha = 0.86$ and normalized such that the total number of continuum photons emitted per baryons between the Ly$\alpha$ line and the Lyman limit is 9690 for Pop II stars and 4800 for Pop III stars \citep{2005ApJ...626....1B}.  $P_{\rm abs}(n,z)$ is the probability that a Ly$n$ photon will be absorbed by the IGM at redshift $z$.  The first $P_{\rm abs}$ factor just accounts for the fact that not all Ly$n$-frequency photons will be absorbed by the IGM, in particular at low redshifts where the neutral hydrogen density is low.  The product of $1-P_{\rm abs}(n',z')$ factors accounts for the idea that a continuum photon that is absorbed at redshift $z$ with frequency $\nu_n$ has to transmit through regions where its frequency is $\nu_{n'}$ (not $\nu_n'$) at redshift $z'$.  And $n_{\rm max}$ is the maximum $n'$ such that a photon with frequency $\nu_{n'}$ at redshift $z'$ will reach redshift $z$ with frequency $\nu_n$, given by
\begin{eqnarray}
n_{\rm max}(z,z') = \left[1-\left(1-\frac{1}{n^2}\right)\left(\frac{1+z'}{1+z}\right)\right]^{-1/2}\, .
\end{eqnarray}
The redshift $z'$ where $n_{\rm max}\to\infty$ is $z_{\rm max}$, which is also the redshift such that a Lyman-limit photon could redshift down to a Ly$n$ photon, given by
\begin{eqnarray}
1+z_{\rm max}(z,n) = \frac{1+z}{1-\frac{1}{n^2}}\, .
\end{eqnarray}
The $z_{\rm max}$ value is important because if $n_{\rm max}\to\infty$, then the product of $1-P_{\rm abs}(n',z')$ in the integrand vanishes because all the entries are less than unity, and an infinite product over values less than unity must tend to zero.  Physically, this just means that the probability of a photon \emph{not} being absorbed while redshifting through an infinite number of Ly$n$ lines is nil.

The absorption probability can be written as $P_{\rm abs}(n,z)=1-\VEV{e^{-\tau}}$, where the average is performed assuming the probability of IGM density perturbations $P(\Delta,z)$ as \citep{2000ApJ...530....1M}
\begin{eqnarray}
\VEV{e^{-\tau}}(z) = \int_0^{100} d\Delta\,P(\Delta,z)e^{-\Delta^2\tau}\, ,
\end{eqnarray}
where $P(\Delta,z)$ is given by
\begin{eqnarray}
P(\Delta,z) = A(z)\exp\left\{-\frac{[\Delta^{2/3}-C_0(z)]^2}{2[2\delta_0(z)/3]^2}\right\}\Delta^{-\beta}\, ,
\end{eqnarray}
and the parameters are given in Table \ref{T:pdf}.  The parameters $\beta$ and $\delta_0$ were fit to simulations for redshifts $2<z<6$ with $\delta_0$ found to follow the function $\delta_0=7.61/(1+z)$.  Since neither $\beta$ nor $\delta_0$ were estimated for redshifts $z<2$, we will not predict continuum emission for these redshifts.  For redshifts $z>6$, we set $\beta=2.5$ and we extrapolate $\delta_0$.  The parameters $A(z)$ and $C_0(z)$ are set by normalizing $P(\Delta,z)$ and $\Delta P(\Delta,z)$ to unity.
\begin{table}[!t]
\begin{center}
\caption{\label{T:pdf} Parameter values for $P(\Delta,z)$, the probability distribution of IGM fluctuations.}
\begin{tabular}{cccc}
\hline
$z$&$A$&$\beta$&$C_0$\\
\hline
2&0.170&2.23&0.623\\
3&0.244&2.35&0.583\\
4&0.322&2.48&0.578\\
6&0.380&2.50&0.868\\
8&0.463&2.50&0.964\\
10&0.560&2.50&0.993\\
12&0.664&2.50&1.001\\
15&0.823&2.50&1.002\\
\hline
\end{tabular}\end{center}
\end{table}

For the optical depth, we use the Gunn-Peterson optical depth, given by \citep{1965ApJ...142.1633G,2006MNRAS.367..259H}
\begin{eqnarray}
\tau_{GP}(n,z) = \frac{3n_H(z)[1-X_e(z)]\lambda_{{\rm Ly}n}^3\gamma_n}{2H(z)}\, ,
\end{eqnarray}
where $\lambda_{{\rm Ly}n}$ is the wavelength of the Ly$n$ line and $\gamma_n$ is the half width at half-maximum (HWHM) of the Ly$n$ resonance, given by
\begin{eqnarray}
\gamma_n = 50\,{\rm MHz}\left(\frac{1-n^{-2}}{0.75}\right)^2\left(\frac{f_n}{0.4162}\right)\, ,
\end{eqnarray}
where $f_n$ is the oscillator strength for the Ly$n$ transition.

\subsection{Intensity Results}
\begin{figure}
\begin{center}
{\scalebox{.5}{\includegraphics{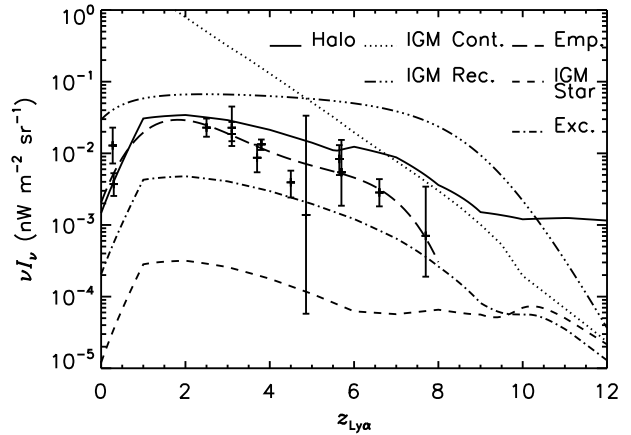}}}
\caption{\label{F:intlya} The Ly$\alpha$ line intensity as a function of $z_{\rm Ly\alpha}=\nu_{\rm Ly\alpha}/\nu_{\rm obs}-1$.  We plot the theoretical emission from dark matter halos (solid), the theoretical emission from the IGM due to stars (short-dashed), the empirical emission (crosses), the cubic fit to the estimated emission from LF measurements (long-dashed).  We see that the halo emission dominates the signal from Ly$\alpha$ emitters and matches the LF measurements.  We also plot the diffuse IGM emission due to recombinations (dot-dot-dot-dashed) and continuum photons (dotted), which dominate at all redshifts.  Finally, we plot the contribution from excitations (dot-dashed) that would contribute to the halo emission.}
\end{center}
\end{figure}

We plot the theoretical intensity from the Ly$\alpha$ emitters (Sec.~\ref{S:halo} and \ref{S:igms}) and the diffuse IGM sources, including recombinations and continuum photons in Fig.~\ref{F:intlya}.  We see from the figure that the halo emission dominates the Ly$\alpha$ emitter emission at all redshifts.  We also plot the empirical predictions of the intensity for $z<8$.  We see that the theoretical halo emission model predicts a slightly higher intensity than the empirical model but close to the errors.  Note that the error bars on the empirical model are larger than true 1-$\sigma$ errors.  Also, the diffuse IGM emission does not conflict with the luminosity function measurements because LFs do not account for diffuse emission.  To better quantify the empirical model, we perform a chi-squared fit on the common logarithm to a quadric function, finding
\begin{eqnarray}
\log_{10}\left(I_{\rm emp}^{\rm Ly\alpha}\right)&=&-2.687+1.568z-0.6995z^2+0.1146z^3\nonumber\\
&&-0.00666z^4\, ,
\end{eqnarray}
in units of ${\rm nW\,m^{-2}\,THz^{-1}\,str^{-1}}$.

In addition, for most of the redshifts in our analysis, the diffuse IGM emission from continuum photons dominates the low-redshift signal, while the halo emission dominates the high-redshift signal, unless the diffuse IGM emission due to recombinations plays a role.  Note that the signal from IGM continuum photons can vary from the fiducial model due to uncertainties in the SFRD and the stellar emissivity at higher redshifts.  The diffuse IGM components provide a strong case for intensity mapping, considering that this probe, contrary to number counts, will access the diffuse IGM emission.  It should be recognized that the model for Ly$\alpha$ emitters at high redshift is indeed a lower limit relevant to the minimum reionization condition; the actual emission from Ly$\alpha$ emitters could be much higher. These results lead us to suspect that an intensity map of Ly$\alpha$ emission will trace Ly$\alpha$ emitters at high redshifts, allowing us to probe LSS in the reionization epoch.

\section{Lyman-alpha line fluctuations} \label{S:modfluc}
In this section we determine the power spectrum for Ly$\alpha$ line fluctuations.  The power spectrum for intensity fluctuations in the Ly$\alpha$-line is given by
\begin{eqnarray}
P^{\rm Ly\alpha}(k,z)=P^{\rm halo}(k,z)+P^{\rm dir}(k,z)+P^{\rm dic}(k,z)\, ,
\end{eqnarray}
where we only include the halo and the diffuse IGM emission because the other sources are subdominant.  Cross-correlations between these sources of emission should be negligible: the diffuse emission fluctuates according to $n_H^2$; assuming a Gaussian distribution of perturbations it should not correlate with the other two sources.

\subsection{Halo Emission}
The power spectrum for the halo emission, and thus the Ly$\alpha$ emitters, is given by
\begin{eqnarray} \label{E:pkhalo}
P^{\rm halo}(k,z)=\nu_{\rm obs}\left[I_{\rm halo}^{\rm Ly\alpha}(z)\right]^2\left[P_{gg}(k,z)+P^{\rm shot}(z)\right]\, ,
\end{eqnarray}
where $P_{gg}(k,z)$ is the galaxy-galaxy clustering power spectrum, and $P^{\rm shot}(z)$ is the shot noise. We implement the halo-model formalism when calculating $P_{gg}(k,z)$, but we use only the two-halo clustering term.  We neglect the one-halo clustering term and nonlinear clustering at small scales since in our case it will always be dominated by the shot noise.  Eq.~\ref{E:pkhalo} models the power spectrum for the theoretical emission model.  For the empirical emission model, we simply replace $I_{\rm halo}^{\rm Ly\alpha}(z)$ with $I_{\rm emp}^{\rm Ly\alpha}(z)$.

\begin{figure}
\begin{center}
{\scalebox{0.5}{\includegraphics{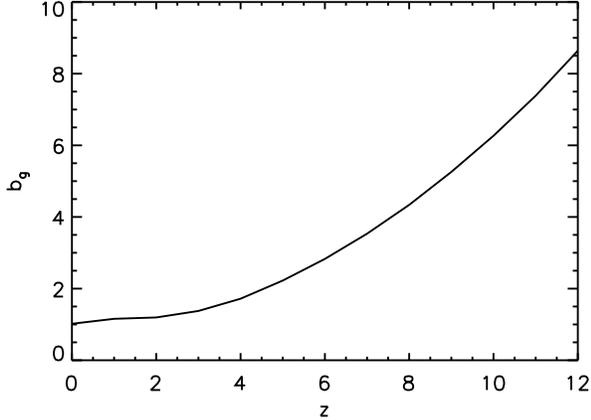}}}
\caption{\label{F:bias} The galaxy clustering bias used in our analysis.}
\end{center}
\end{figure}

We use a galaxy-galaxy power spectrum similar to the power spectrum discussed in \citet{2002PhR...372....1C}.  The formulas are given by
\begin{eqnarray}\label{E:pgg}
P_{gg}(k,z)&=&b_g^2(z)P_{\rm lin}(k,z)\nonumber\\
b_g&=&\frac{\int_{M_{\rm min}}^{M_{\rm max}} dM\,\VEV{N_g(M)}n(M)b(M,z)}{\int_{M_{\rm min}}^{M_{\rm max}} dM\,\VEV{N_g(M)}n(M)}\, ,
\end{eqnarray}
where $b(M,z)$ is the halo bias, $\VEV{N_g(M)}$ is the mean number of galaxies in a halo of mass $M$, and $b_g$ becomes the galaxy clustering bias, shown in Fig.~\ref{F:bias}.  For the shot noise, we use the expression from dark matter, given by
\begin{eqnarray}
P^{\rm shot}(z)&=&\int_{M_{\rm min}}^{M_{\rm max}} dM\,n(M)\left[\frac{M}{\rho_mf_{\rm coll}}\right]^2\, .
\end{eqnarray}
In our calculation we use the entire mass range; however, the higher end of this mass range will be probed by number count surveys like JWST, meaning that any new science from the shot noise in an intensity mapping survey would be captured by a high-mass cutoff corresponding to the minimum flux of a competing number counts survey.  This will not affect the shot noise science from the diffuse IGM because it is inaccessible to number count surveys.

\subsection{Diffuse IGM from recombinations}
The power spectra for the diffuse IGM components are more complicated than the contributions from halo and stellar IGM emission.  For $P^{\rm dir}(k,z)$, the intensity scales as $n_H^2$ times factors of $X_e$.  We neglect the fluctuations in the ionization fraction in this derivation, which may underestimate the correlation signal.  This choice makes the intensity fluctuations of the form
\begin{eqnarray}
\delta I_{\rm diff\,rec/coll}^{\rm Ly\alpha}(z,\hatn) &=& I_{\rm diff\,rec/coll}^{\rm Ly\alpha}(z)\nonumber\\
&&\times\left\{\frac{n_H^2[\hatn\chi(z)]-\overline{n_H^2}(z)}{\overline{n_H^2}(z)}\right\}\, .
\end{eqnarray}
The formula for $P^{\rm dir}(k,z)$ will be similar to Eq.~\ref{E:pkhalo}, except that we require the 3D power spectrum for $n_H^2$, $P_{H^2}(k)$.  We skip the derivation here and just present the result, where
\begin{eqnarray}
P_{H^2}(k)&=&\frac{1}{2\pi^2}\int dk'\,k'^2P_H(k')\nonumber\\
&&\times\int_{-1}^{1}d\mu\,P_H(\sqrt{k^2+k'^2-2kk'\mu})\, ,
\end{eqnarray}
where $P_H(k)$ is the two-point power spectrum for hydrogen atoms (or baryons).  The fluctuations of the baryons do not trace fluctuations in the dark matter, particularly on small scales where baryonic interactions smooth out the perturbations.  We chose a toy model for the baryonic power spectrum used by \citet{2005ApJ...630..643M}, where $P_{H}(k)=F_b(k/k_F)P_{gg}(k,z)$, where $k_F(z)=34[\Omega_m(z)]^{1/2}h$ Mpc$^{-1}$ \citep{1998MNRAS.296...44G} filters out small scale perturbations and
\begin{eqnarray}
F_b(x)=\frac{1}{2}\left[e^{-x^2}+\frac{1}{(1+4x^2)^{1/4}}\right]\, .
\end{eqnarray}
for $z<8$ and $F(x)=1$ for $z>8$.

The shot noise component for $C_\ell^{dir}$ requires a four-point calculation of areal number densities.  We do this for a Poisson distribution and find that the shot noise component for the correlation of a wavemode is $C(\veck_i)=Q(\bar{N})/\bar{n}$, where $\bar{n}$ is the average number density of atoms in the IGM and $\bar{N}$ is the average number of IGM sources per pixel, both of which will depend on the minimum luminosity detectable by the spectrograph while $\bar{N}$ also depends on the pixelation scheme, and $Q(\bar{N})$ is given by
\begin{eqnarray}
Q(\bar{N})=\frac{4\bar{N}^2+6\bar{N}+1}{(\bar{N}+1)^2}\, .
\end{eqnarray}
$Q(\bar{N})$ only has values between 1 for $\bar{N}\ll1$ and 4 for $\bar{N}\gg1$.  We decide to set $Q=2$, knowing that we could misestimate the shot noise by half an order of magnitude.  Also, although this was calculated for only Poisson distribution, we apply this formula to the Poisson-like distribution seen in halo models by setting the shot noise equal to $P_{H^2}^{\rm shot}(z)=Q(\bar{N})P^{\rm shot}(z)\simeq2P^{\rm shot}(z)$.

\subsection{Diffuse IGM from continuum photons}
The power spectrum for diffuse IGM emission due to the stellar continuum is complicated by having two sources of fluctuations: the local IGM and the stars at higher redshifts that source the emission.  The local IGM power spectrum elucidates the environment at a particular redshift, while the power spectrum from the higher-redshift stars mixes perturbations from redshifts greater than the redshift of the signal but less than $z_{\rm max}(z,n=2)$, the maximum redshift that a Lyman-limit line can be redshifted to a Ly$\alpha$ line at redshift $z$.  In this analysis we will consider both power spectra.

For the local IGM, the perturbations come from the factor $P_{\rm abs}(n,z)$ in Eq.~\ref{E:Icont}, which contains perturbations in the local gas density.  The absorption probability at a particular location is given by $P_{\rm abs}(\vecx,n,z)=1-e^{-\Delta^2(\vecx)\tau}$, which is not linear in $\Delta^2$.  Thus we take the derivative to get the intensity perturbation, giving us
\begin{eqnarray}
\delta[I_{\rm diff\,cont}^{\rm Ly\alpha}](\vecx) &=& \frac{h_P\nu_{\rm Ly\alpha}}{4\pi}\sum_{n=2}^{\infty}f_{\rm rec}(n)\int_z^\infty dz'\,\frac{c}{(1+z')H(z')}\nonumber\\
&&\times\dot{\rho_*}(z')\epsilon(\nu_n') \VEV{\tau e^{-\tau}}(z)\delta_{H^2}(\vecx)\nonumber\\
&&\times\prod_{n'=n+1}^{n_{\rm max}}[1-P_{\rm abs}(n',z')]\, ,
\end{eqnarray}
where
\begin{eqnarray}
\VEV{\tau e^{-\tau}}(z) = \int_0^{100} d\Delta\,P(\Delta,z)\Delta^2\tau e^{-\Delta^2\tau}\, .
\end{eqnarray}
Therefore, we can write the power spectrum for the local IGM as
\begin{eqnarray}
P^{\rm dic}_{\rm local}(k,z)=\left[\nu_{\rm obs}J_{\rm cont}^{\rm local}(z)\right]^2\left[P_{H^2}(k,z)+P_{H^2}^{\rm shot}(z)\right]\, ,
\end{eqnarray}
where
\begin{eqnarray}
J_{\rm cont}^{\rm local}(z) &=& \frac{h_P\nu_{\rm Ly\alpha}}{4\pi}\sum_{n=2}^{\infty}f_{\rm rec}(n)\int_z^\infty dz'\,\frac{c}{(1+z')H(z')}\nonumber\\
&&\times\dot{\rho_*}(z')\epsilon(\nu_n') \VEV{\tau e^{-\tau}}(z)\nonumber\\
&&\times\prod_{n'=n+1}^{n_{\rm max}}[1-P_{\rm abs}(n',z')]\, .
\end{eqnarray}

For the higher-redshift stars, we rewrite the measured intensity (Eq.~\ref{E:Icont}) as
\begin{eqnarray}
I_{\rm diff\,cont}^{\rm Ly\alpha}=\frac{c}{4\pi}\int_z^\infty dz'\frac{\dot{\rho_*}(z')A_{\rm cont}(z,z')}{H(z')(1+z')}\, ,
\end{eqnarray}
where $A_{\rm cont}(z,z')$ is given by
\begin{eqnarray}
A_{\rm cont}(z,z')&=&h_P\nu_{\rm Ly\alpha}\sum_{n=2}^{\infty}f_{\rm rec}(n)\epsilon(\nu_n') P_{\rm abs}(n,z)\nonumber\\
&&\times\prod_{n'=n+1}^{n_{\rm max}}[1-P_{\rm abs}(n',z')]\, .
\end{eqnarray}
In this form, we can easily Fourier transform the SFRD, and assuming it traces the galaxy perturbations, we can write a contribution to the foreground power spectrum as
\begin{eqnarray}
P^{\rm dic}_{\rm fg}(k,z)=[\nu J_{\rm cont}^{\rm sfrd}]^2P_{gg}(k)+P_{\rm dic}^{\rm shot}\, ,
\end{eqnarray}
where $P_{gg}(k) = P_{gg}(k,z=0)$,
\begin{eqnarray}
\nu J_{\rm cont}^{\rm sfrd}&=&\frac{c\nu_{\rm obs}(z)}{4\pi}\int_z^\infty dz'\frac{\dot{\rho_*}(z')A_{\rm cont}(z,z')}{H(z')(1+z')}\nonumber\\
&&\times b_g(z')\left[\frac{D(z')}{D(0)}\right]\, ,
\end{eqnarray}
$D(z)$ is the growth function, and
\begin{eqnarray}
P_{\rm dic}^{\rm shot} &=& \left(\frac{c\nu_{\rm obs}(z)}{4\pi}\right)^2\int_z^\infty dz'\left[\frac{\dot{\rho_*}(z')A_{\rm cont}(z,z')}{H(z')(1+z')}\right]^2\nonumber\\
&&\times P^{\rm shot}(z')\, .
\end{eqnarray}

\subsection{Results}
\begin{figure}
\begin{center}
{\scalebox{.4}{\includegraphics{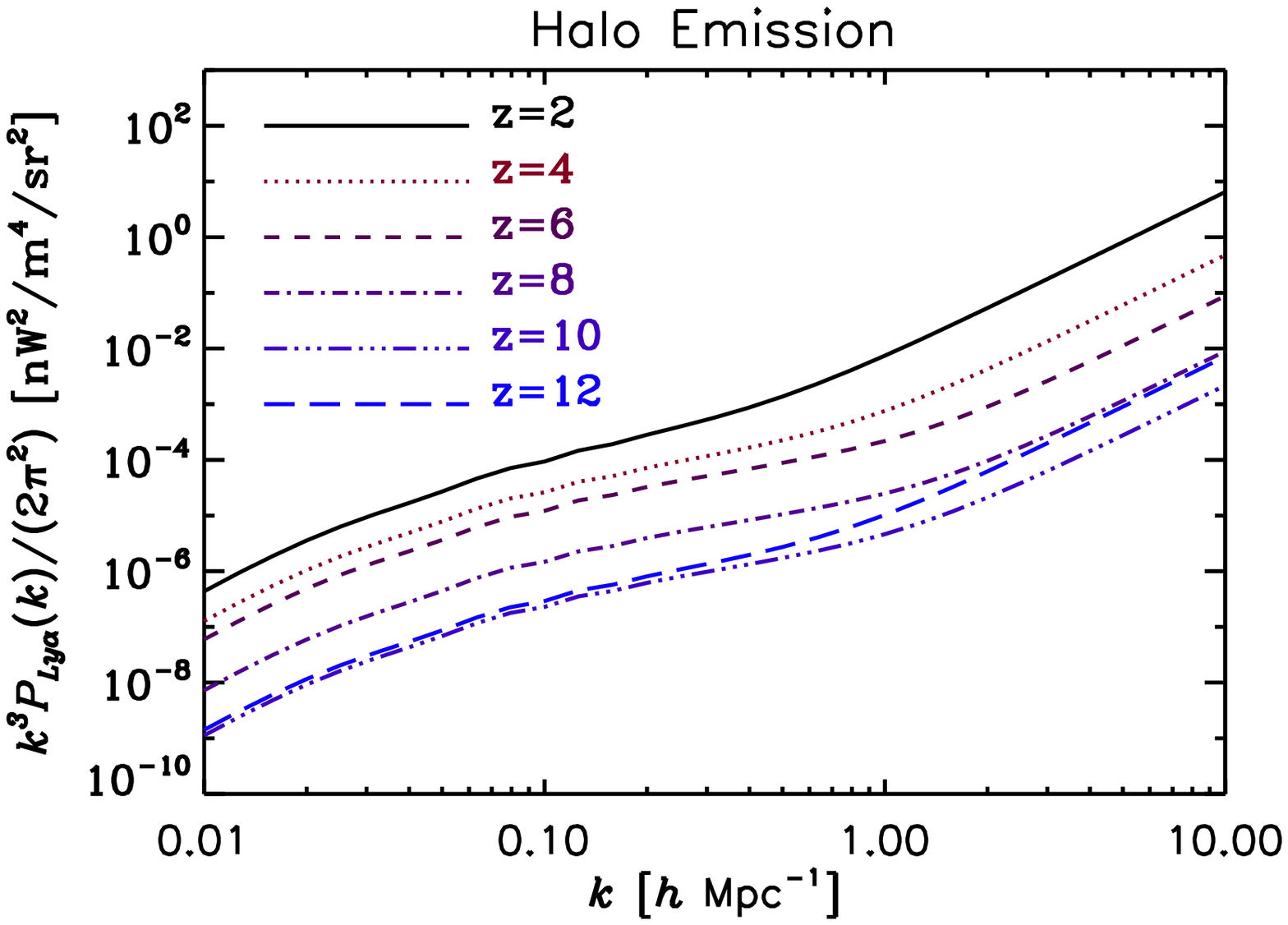}}}
{\scalebox{.4}{\includegraphics{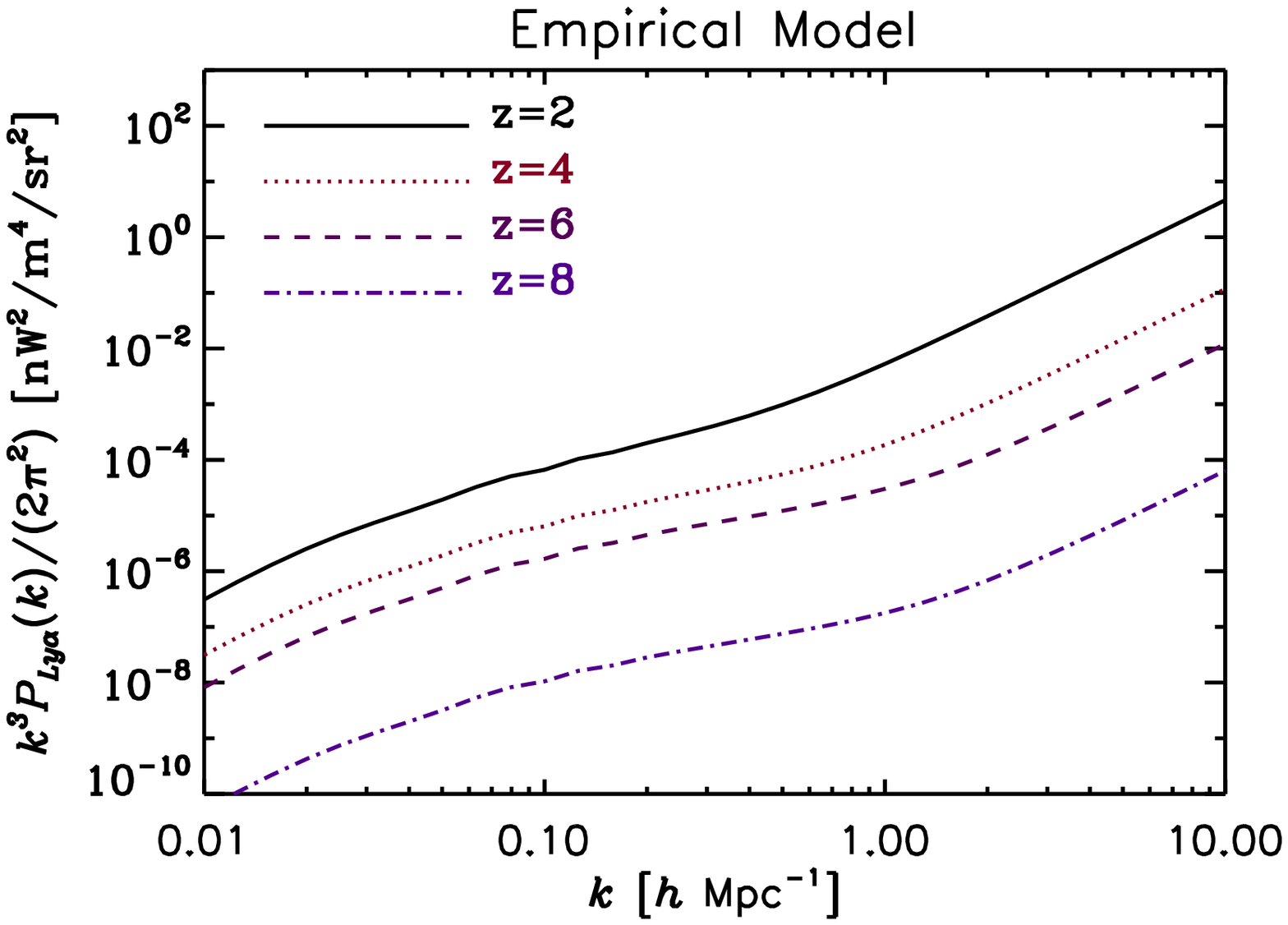}}}
\caption{\label{F:pklyahalo} The theoretical power spectra for halo emission (top), as well as the empirical model (bottom), including both the clustering and shot noise signals.}
\end{center}
\end{figure}

\begin{figure}
\begin{center}
{\scalebox{.4}{\includegraphics{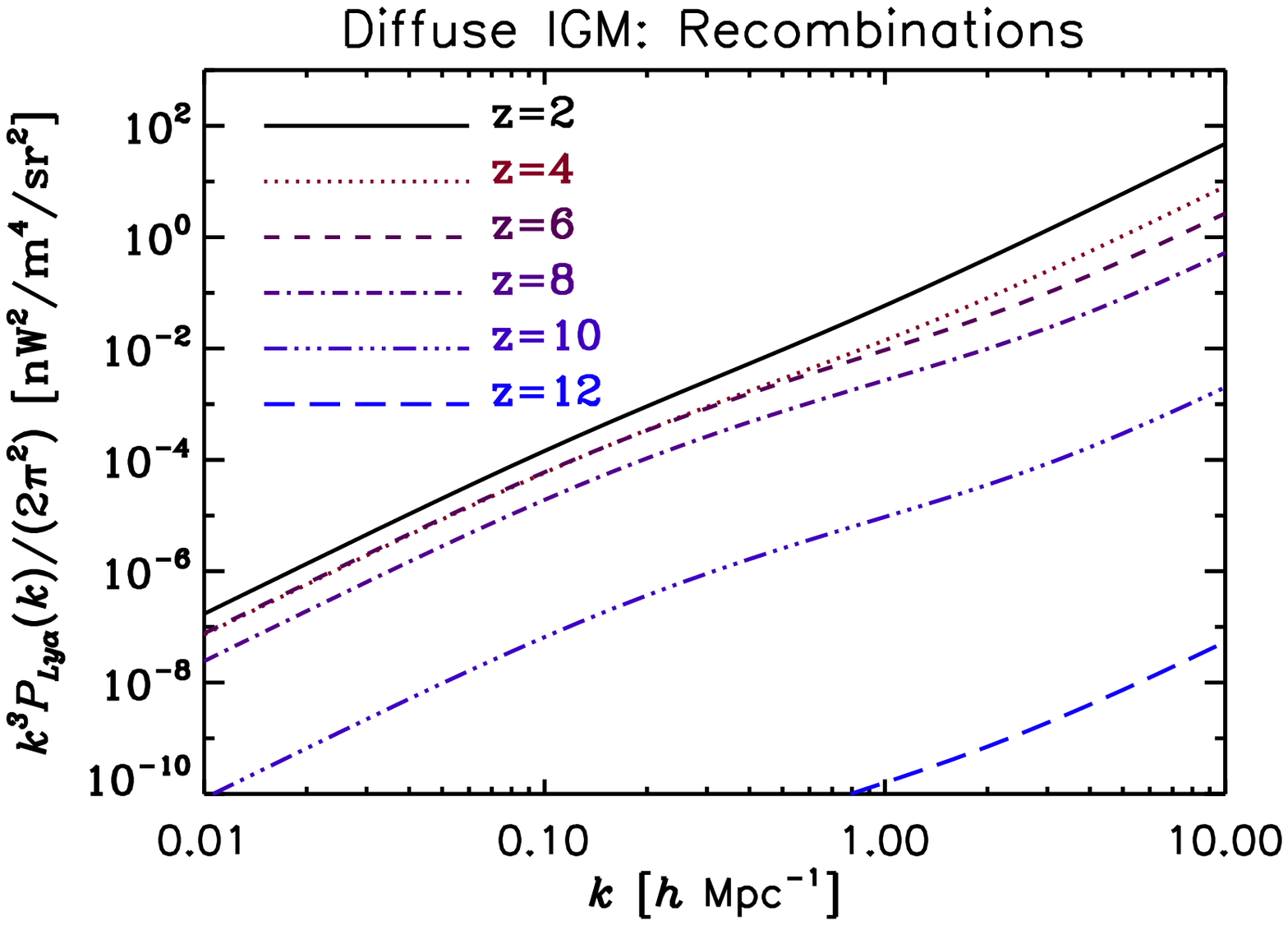}}}
{\scalebox{.4}{\includegraphics{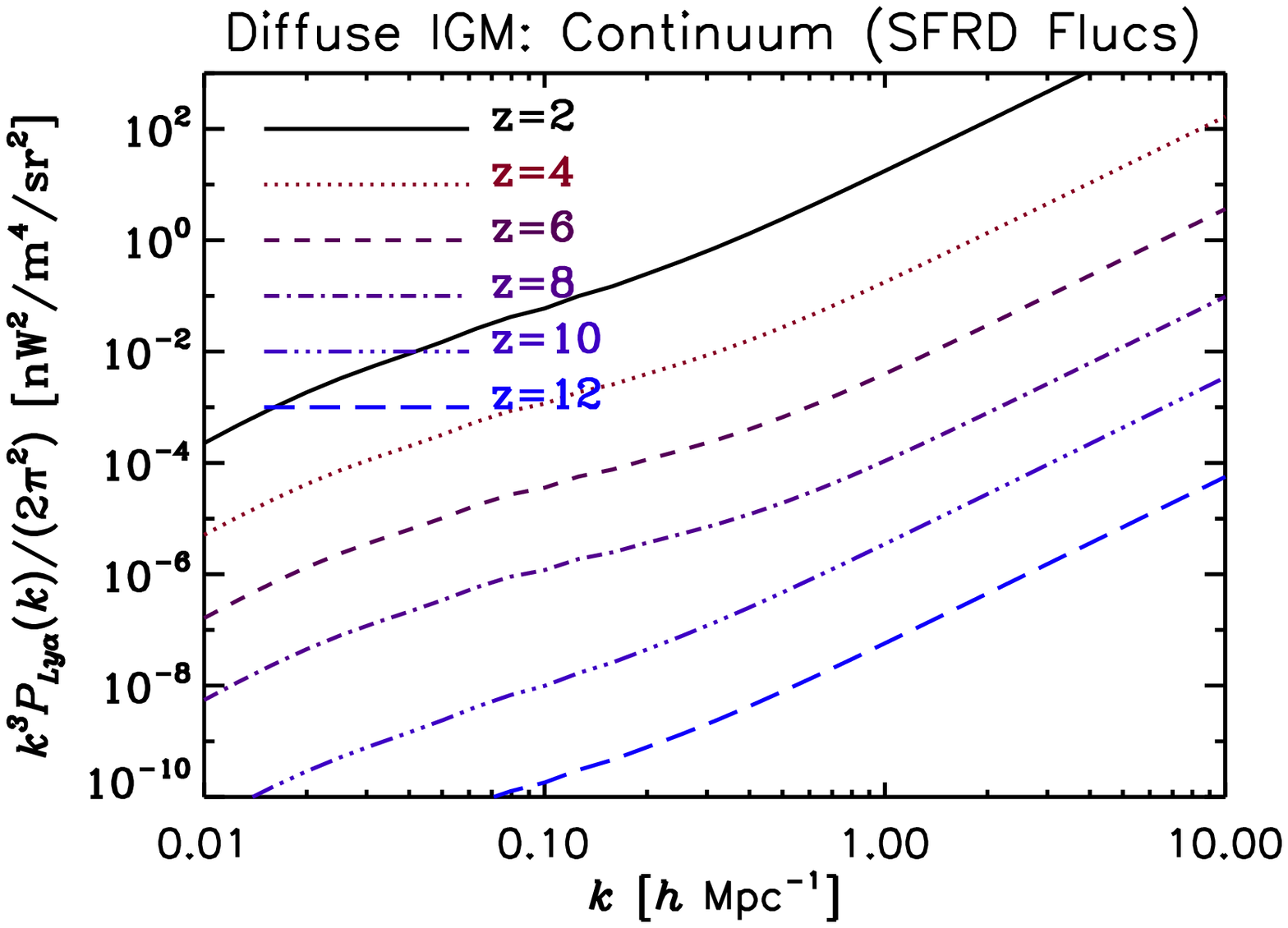}}}
{\scalebox{.4}{\includegraphics{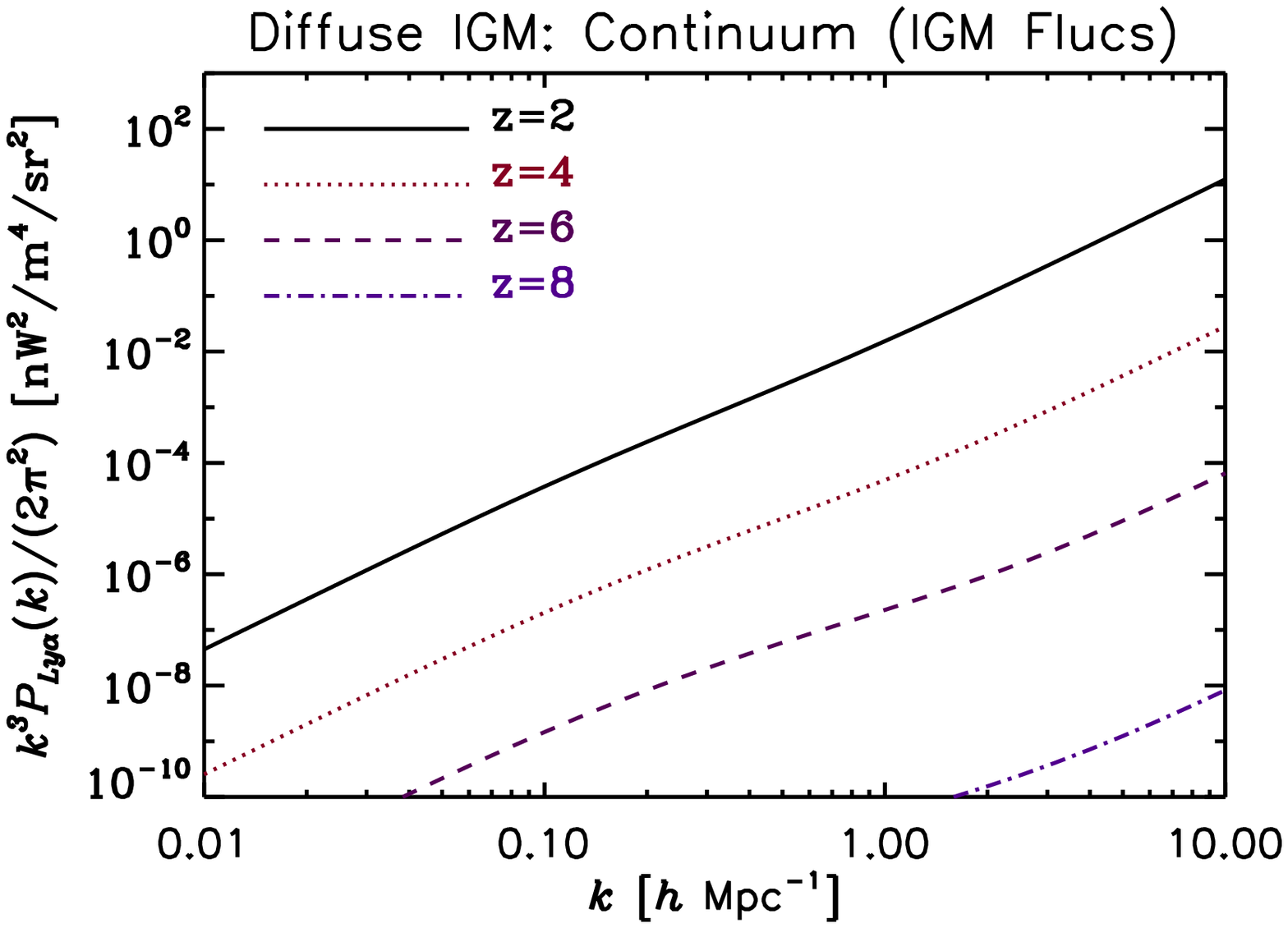}}}
\caption{\label{F:pklyadiff} The power spectra for diffuse IGM emission from recombinations (top) and continuum photons with SFRD fluctuations (center) and local IGM (bottom), including both the clustering and shot noise signals.}
\end{center}
\end{figure}

\begin{figure}
\begin{center}
{\scalebox{.4}{\includegraphics{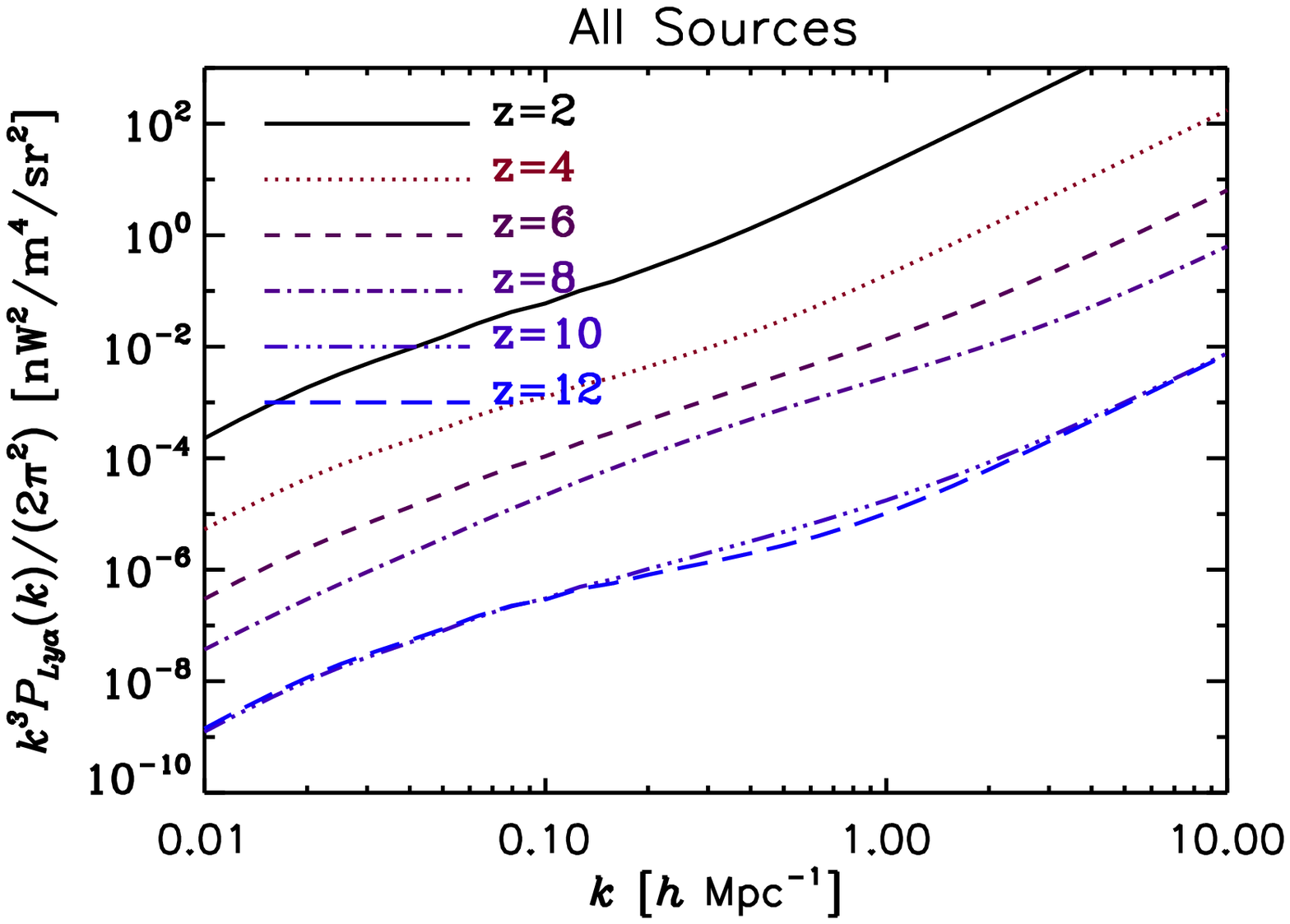}}}
{\scalebox{.4}{\includegraphics{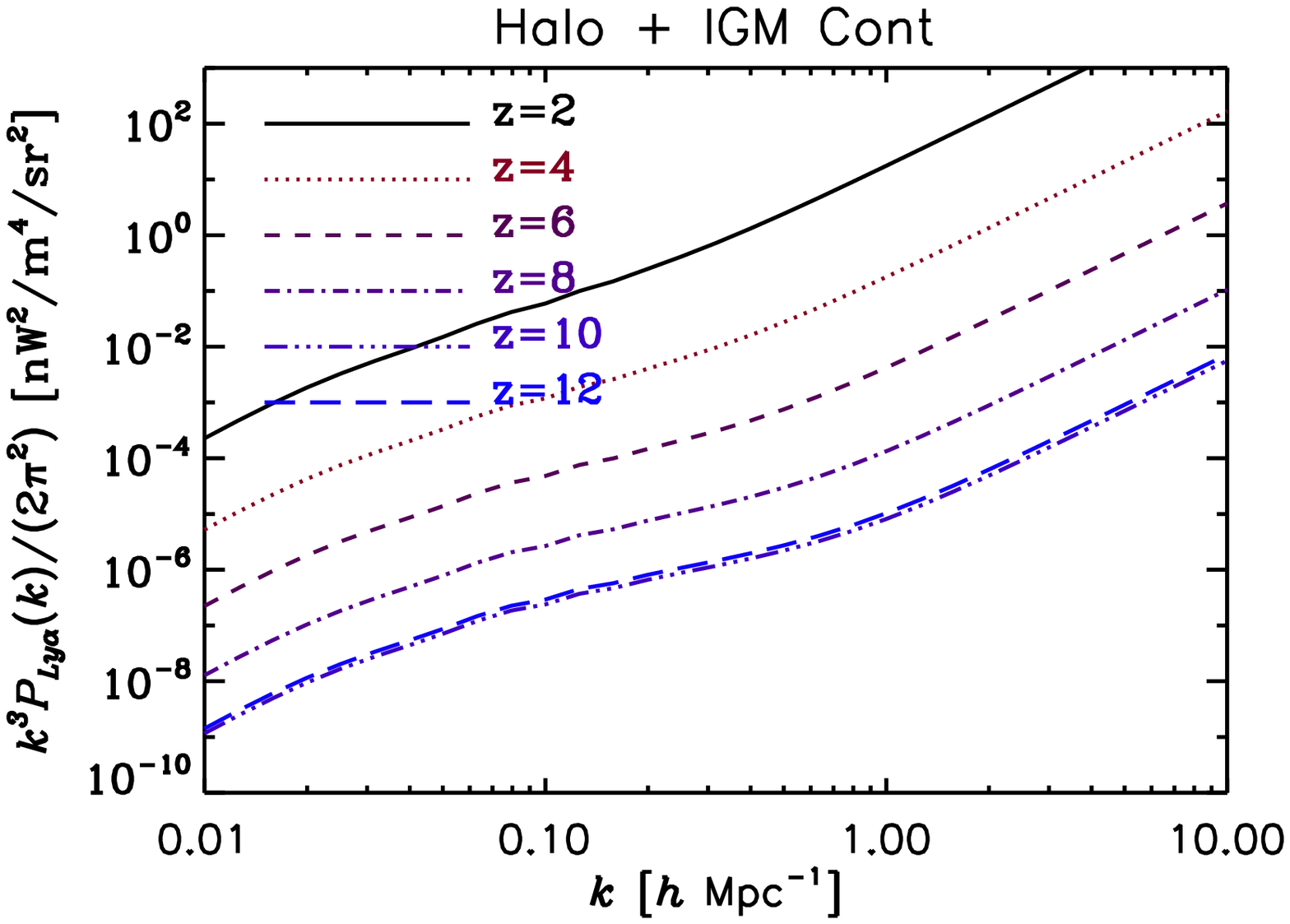}}}
\caption{\label{F:pklyaall} The total power spectra for Ly$\alpha$ emission including various sources.  The first panel includes halo emission and diffuse IGM emission from recombinations and continuum photons, while the bottom panel does not include diffuse IGM recombinations.}
\end{center}
\end{figure}

We plot the 3D power spectra for Ly$\alpha$ emission for the various sources, as well as the empirical model, at various redshifts in Fig.~\ref{F:pklyahalo}.  In Fig.~\ref{F:pklyadiff}, we plot the power spectra for both diffuse IGM components, and in Fig.~\ref{F:pklyaall} we plot combinations.  Since the diffuse IGM emission due to recombinations is more speculative, we plot in Fig.~\ref{F:pklyaall} both the sum of all the sources and the sum of just the halo and the diffuse IGM due to continuum photons (no recombinations).  \emph{We will consider the latter sum as our fiducial model for the remainder of this work.}  According to our fiducial model, an auto-power spectrum measurement of Ly$\alpha$ emission will characterize at low redshifts the diffuse IGM emission from higher-redshift stars through their continuum photons.  At intermediate redshifts during late reionization this diffuse IGM emission will be comparable to the emission from Ly$\alpha$ emitters.  At high redshifts ($z\gtrsim10$), Ly$\alpha$ emitters will be characterized by the Ly$\alpha$ power spectrum.  Another possible avenue is through cross-correlations with high-redshift galaxy tracers (e.g. CO, CII, 21cm).  This approach could work at small scales because the power spectrum of IGM emission from continuum photons with SFRD fluctuations, the competitor of Ly$\alpha$ emitters, are sourced by fluctuations from redshifts higher than the emission redshift, and this would drop out of a cross-correlation.  The power spectrum of IGM emission from continuum photons, but due to IGM fluctuations, could correlate with galaxies on small scales due to nonlinearities (since it would be a 3-point correlation), but it should be much smaller than the halo emission signal.  Thus, we expect Ly$\alpha$ emitters and IGM perturbations to be probed using Ly$\alpha$ emission.

Of course, this picture depends on the various fiducial parameters.  The SFRD can change both the halo and diffuse Ly$\alpha$ IGM emission from continuum photons.  The halo emission also has uncertainties in the dust model, with recent measurements suggesting less obscuration that we assume \citep{2013arXiv1305.3613D}.  The IGM Ly$\alpha$ emission from continuum photons can also vary due to the emissivity.  And the emission from diffuse IGM due to recombinations can vary over an order of magnitude based on the IGM gas temperature.  Also, a change in the clustering bias can greatly impact the power spectrum signal.  Finally, it has been shown that fluctuations in the transmission environment of Ly$\alpha$ photons can cause redshift-space distortions, affecting the clustering power spectrum apart from the intensity \citep{2011MNRAS.415.3929W,2013MNRAS.431.1777G}.  More measurements from various probes, including intensity mapping studies and photometry studies are definitely needed to characterize Ly$\alpha$ emitters and the IGM.

\section{Forecasts} \label{S:forecasts}
We assess the ability of near-IR observatories to measure Ly$\alpha$ line fluctuations.  The main sources to the statistical errors in the Ly$\alpha$ measurements are cosmic variance, instrumental errors, and shot noise.  Also, low-redshift galaxies can emit in the near-IR and supersede the Ly$\alpha$ signal.  Of course, it is impossible to remove all low-redshift galaxies.  However, by removing by hand all low-redshift galaxies with apparent AB magnitudes less than a certain value $m_{\rm lim}$, we can detect Ly$\alpha$ fluctuations if the angular power spectrum for any remaining low-$z$ galaxies at a particular wavelength becomes significantly less than the spectrum for Ly$\alpha$ line fluctuations at the same wavelength.

\subsection{Instrumental sensitivity} \label{S:instrum}
\begin{figure}
\begin{center}
{\scalebox{.4}{\includegraphics{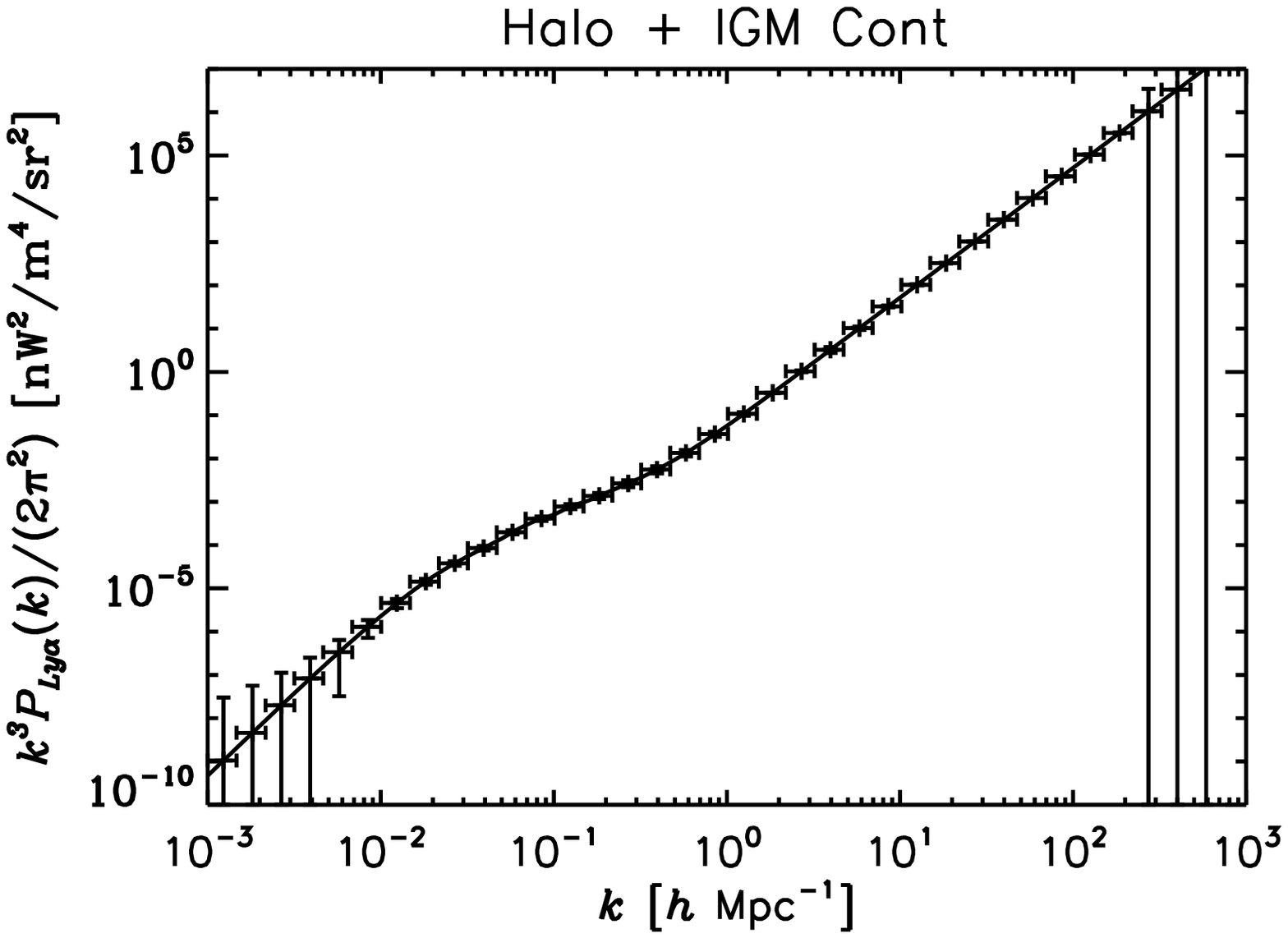}}}
{\scalebox{.4}{\includegraphics{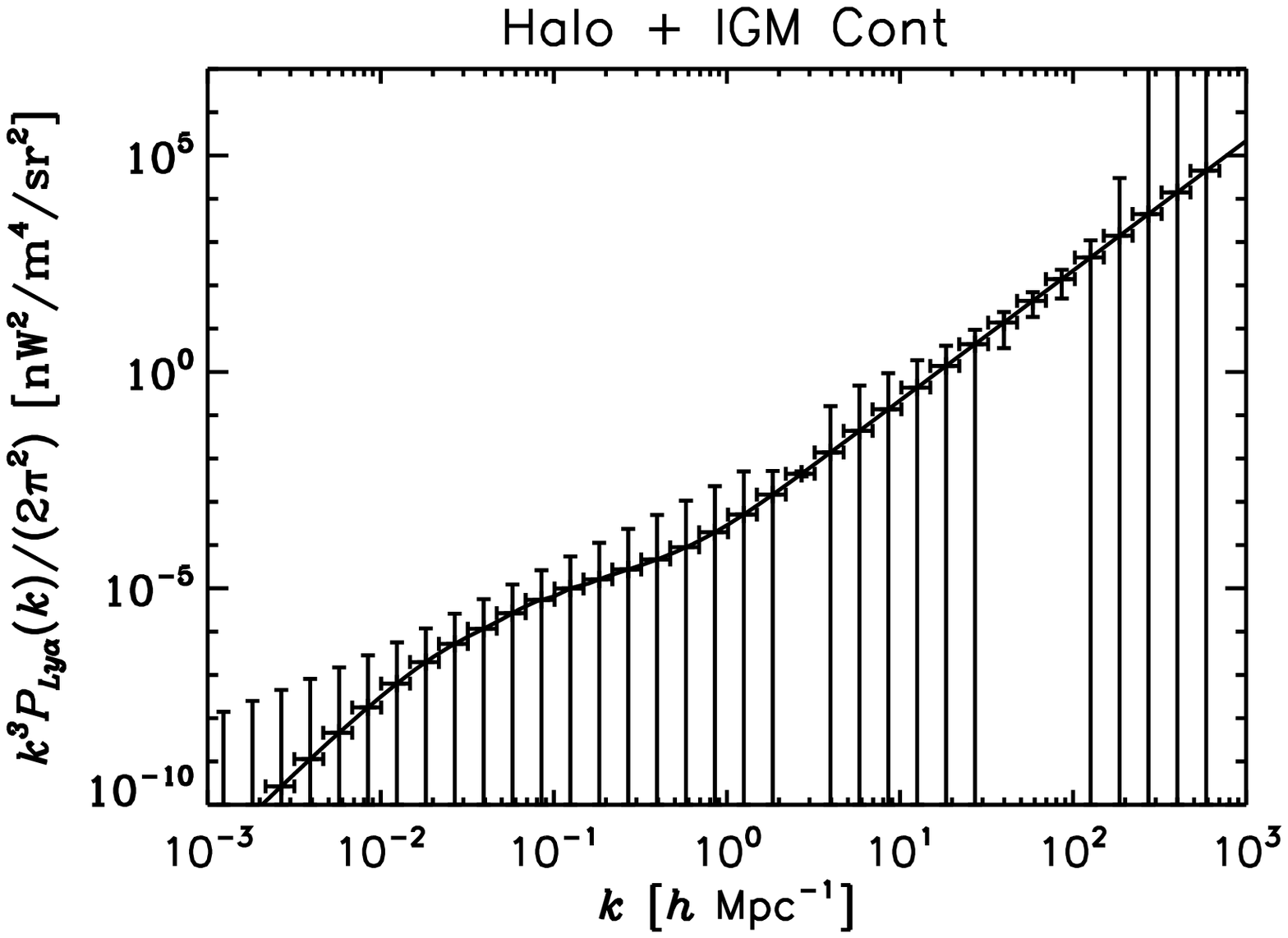}}}
\caption{\label{F:pksig} The total power spectra for Ly$\alpha$ emission with instrumental uncertainties.  The top (bottom) panel plots the power spectrum with 1$\sigma$ errors over the redshift range $z=4-5$ ($z=7-8$).  Note that cosmic variance is not included in the errors.}
\end{center}
\end{figure}

We begin by determining uncertainties in the 3D power spectrum measurement due to instrumental sensitivity.  For our analysis, we consider a configuration for an instrument dedicated to measuring Ly$\alpha$ fluctuations across cosmic time.  We chose two spectral bands with ranges of (0.5-0.9$\mu$m) and (0.9-1.6$\mu$m), corresponding  to pre-EoR and the EoR, and noise values shown in Table \ref{T:inst}.  We give our fiducial instrument a 6 arcsec beam full width at half-maximum (FWHM) and a spectrometer with a resolution $R=40$.  We choose a survey over 200 deg$^2$.  As an illustration we plot 3D power spectra for redshift ranges $z=4-5$ and $7-8$ in Fig.~\ref{F:pksig}.
\begin{table}
\begin{center}
\caption{\label{T:inst} The sensitivity $\sigma_N$ of our fiducial instrument in units of ${\rm (nW\,m^{-2}\,s^{-1})^2}$ for a 1 arcmin$^2$ pixel and the resulting SNR for each band.  We include SNR values for both the fiducial model, the halo emission only, and the fiducial model plus diffuse IGM emission from recombinations.  In parentheses we list the values when cosmic variance is included.}
\begin{tabular}{c|cc}
\hline
Band&0.5-0.9$\mu$m&0.9-1.6$\mu$m\\
\hline
$\sigma_N$&0.5&0.3\\
Fiducial&8400 (5900)&44 (42)\\
Halo only&17 (17)&3.6 (3.5)\\
Fig+IGM recomb&8600 (6000)&220 (170)\\
\hline
\end{tabular}\end{center}
\end{table}

\begin{figure}
\begin{center}
{\scalebox{.5}{\includegraphics[width=\textwidth]{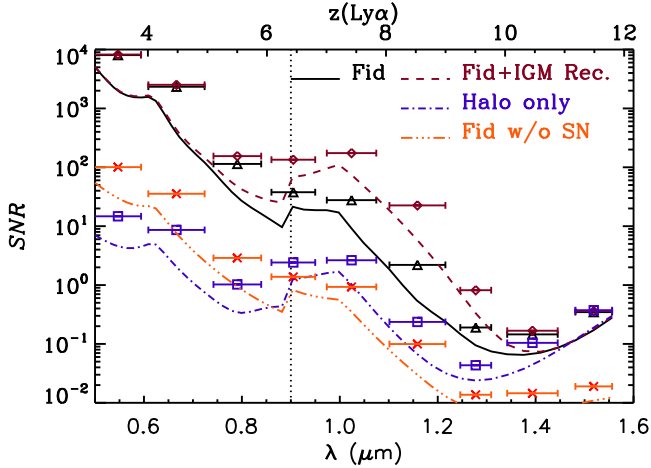}}}
\caption{\label{F:snrz} Forecasts for SNR of the Ly$\alpha$ angular power spectrum in spectral bins for various instrumental configurations.  The spectral bin widths are determined by the spectral resolution, which is $R=40$ for our hypothetical survey.  The bars denote the SNR over redshift ranges $\Delta z=1$.  Note that we include SNRs for the fiducial model (solid, triangles), the halo emission only (dot-dashed, squares), and the fiducial model plus diffuse IGM emission from recombinations (dashed, diamonds). We also include the fiducial model with the shot noise removed from the signal (dot-dot-dot-dashed, X).  The dotted line divides the first and second spectral bands.}
\end{center}
\end{figure}

In Fig.~\ref{F:snrz} we plot the signal-to-noise ratio SNR(z) for the Ly$\alpha$ power spectrum, given by
\begin{eqnarray}
SNR^2(z)&=&\frac{V_{\rm survey}(z)}{4\pi^2}\int_0^1 d\mu\int_{\rm k_{\rm min}}^{\rm k_{\rm max}} dk\,k^2\nonumber\\
&&\times\left[\frac{P^{\rm Ly\alpha}(k,z)}{P_N(z)W(k,\mu,z)}\right]^2\, ,\nonumber\\
\end{eqnarray}
where $V_{\rm survey}(z)$ is survey volume at redshift bin $z$, $P_N(z)=\sigma_N^2V_{\rm pix}/\Omega_{\rm pix}$, the square of the instrumental sensitivity times the pixel volume divided by the pixel solid angle, $\mu=\cos\theta$ in $k$-space with the integration over the upper half-plane, and $W(k,\mu,z)$ is the window function given in \citet{2011ApJ...741...70L} as
\begin{eqnarray}
W(k,\mu,z) = e^{(\mu k/k_{\parallel,res})^2+(1-\mu^2)(k/k_{\perp,res})^2}\, ,
\end{eqnarray}
and $k_{\parallel,res}=R\,H(z)/[c(1+z)]$ and $k_{\perp,res}=2\pi/(\chi(z)\sigma_b)$ are the wavenumber resolutions in directions parallel and perpendicular to the line of sight and $\sigma_b = 0.4245$ FWHM.  We do not include cosmic variance in our expression for $SNR$ because we are just interested in detection for this calculation.  However, cosmic variance must be included when constraining background cosmology and we list results for band $SNR$ including cosmic variance in Table \ref{T:inst}.  Note that we set $k_{\rm min}=0.002\,h$/Mpc and $k_{\rm max}=10^3\,h$/Mpc, where $k_{\rm min}$ is limited by the survey volume and $k_{\rm max}$ is limited only by the instrumental beam size.

We also calculate the signal-to-noise ratio for $P(k)$ detection in the two bands, given by
\begin{eqnarray}
{\rm SNR}^2 =\sum_i{\rm SNR}^2(z_i)\, ,
\end{eqnarray}
where $z_i$ runs over all the redshift with corresponding spectral bins within each band, and we do the same for redshift ranges with size $\Delta z=1$.  The values for the small redshift ranges are plotted in Fig.~\ref{F:snrz}, and we list the values for the two bands in Table \ref{T:inst}.  We see from this table that Ly$\alpha$ fluctuations for $z<8$ should be well probed by this deep survey, possibly making the very end of reionization able to be probed.  The signal is mainly dominated by shot noise.  If the shot noise is removed, only the clustering signal from redshifts $z<5$ could possibly be detected, although this is an underestimate since nonlinear clustering effects were neglected.  It also appears the fluctuations at $z<5$ from just the Ly$\alpha$ emitters can also be detected, possibly through a cross-correlation with another LSS tracer.  But most of the Ly$\alpha$ emitter signal is from shot noise which becomes undetectable when objects with masses detectable by JWST are removed.  Also, while the fluctuations deeper within reionization can be detected over the entire Band 2 with $SNR\sim44$, a tomography study in the reionization region is beyond reach for $z>8$.  At redshifts $z>5$, the signal is noise-limited.  Increasing the sensitivity by a factor of 3 would allow fluctuations from $z<9$ to be detected, but without incredibly long integration times, this is well beyond these specifications. One solution may be to do a very deep survey over 20 deg$^2$.  In this case, keeping the total survey integration time constant, the SNR in the second band increases to 95, which can provide much better tomography.  The prospects do improve in the mid- to high-redshift range if diffuse IGM emission due to recombinations plays a role, but this is highly speculative.  Changes in the various model parameters (e.g. $M_{\rm min}$, $b_g$, $T_{\rm gas}^{\rm halo}$, $SFRD$, $X_e$) can change this picture somewhat.  A significant uncertainty is due to the high-redshift intensity being set to the minimum required level to sustain reionization.  A significant increase in this signal could very well place the power spectrum from the EoR within the range for tomography.

\subsection{Theoretical Continuum Foregrounds from stars}
In this section we calculate the foreground power spectrum of all continuum photons similar to the methodology presented in \citet{2010ApJ...710.1089F}, including stellar, free-free, free-bound, and two-photon emission.  The power spectrum for the continuum sources at observed frequency $\nu_{\rm obs}$ is given by
\begin{eqnarray}\label{E:powcont}
P^{\rm cont}(k) = [\nu_{\rm obs} J_{\rm cont}^{\rm fg}]^2P_{gg}(k)+P_{\rm cont}^{\rm shot}\, .
\end{eqnarray}
The clustering term is written as
\begin{eqnarray}
J_{\rm cont}^{\rm fg} = \frac{c}{4\pi}\int_z^\infty dz'\frac{p[\nu_{\rm obs}(1+z'),z']}{H(z')(1+z')}b_g(z')\left[\frac{D(z')}{D(0)}\right]\, ,\nonumber\\
\end{eqnarray}
where $p(\nu,z)$ is the total emissivity of continuum photons, and the shot noise is given by
\begin{eqnarray}
P_{\rm cont}^{\rm shot} = \left(\frac{c}{4\pi}\right)^2\int_z^\infty dz'\left\{\frac{p[\nu(1+z),z']}{H(z')(1+z')}\right\}^2P^{\rm shot}(z')\, .\nonumber\\
\end{eqnarray}
The emissivity can be broken into the various emission sources as
\begin{eqnarray}
p(\nu,z) &=& p^*(\nu,z)+(1-f_{\rm esc})[p^{ff}(\nu,z) \\
&+&p^{fb}(\nu,z)+p^{2\gamma}(\nu,z)]\, ,
\end{eqnarray}
where $p^*(\nu,z)$, $p^{ff}(\nu,z)$, $p^{fb}(\nu,z)$, and $p^{2\gamma}(\nu,z)$ are the emissivities for stellar, free-free, free-bound, and two-photon emission, respectively.  The  individual emissivities are in the form of Eq.~\ref{E:emis} with $L_\nu^\alpha(m)$ for each emission mechanism presented in \citet{2006ApJ...646..703F}.  We also include the effect of galactic extinction by using the Calzetti extinction law \citep{2000ApJ...533..682C} assuming an average extinction of $E(B-V)=0.2$.

\begin{figure}
\begin{center}
{\scalebox{.5}{\includegraphics{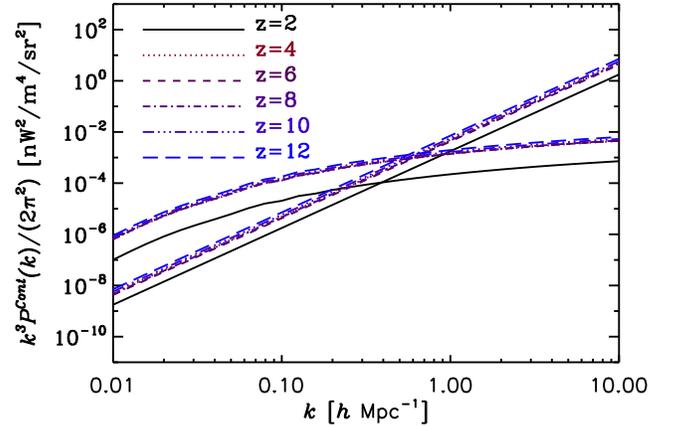}}}
\caption{\label{F:powcont} The 3D auto-power spectra of the continuum emission at various frequencies with emission from source redshifts $z<2$ removed.  In the plot, $z=\nu/\nu_{\rm Ly\alpha}-1$, not the emission source redshift.  The curved lines are the clustering spectrum, while the straight lines are the shot noise.  We assume a Chabrier stellar halo IMF with Pop II and Pop III stars and a minimum halo mass $M_{\rm min}=10^{9}h^{-1}M_\odot$.  We see that our fiducial power spectrum should overtake the continuum spectrum at low redshifts but will be subdominant at EoR redshifts.}
\end{center}
\end{figure}

We present the continuum power spectrum for various frequencies in Fig.~\ref{F:powcont}.  We neglect contributions from stars with $z<2$ because in this calculation we assume we can cull these by hand.  We find that at lower Ly$\alpha$ redshifts, our fiducial power spectrum is greater than the continuum emission power spectrum.  However, at redshifts within the EoR ($z\gtrsim6$), the continuum clearly dominates over our Ly$\alpha$ power spectrum.  

\subsection{Line Foregrounds from galaxies}

It will still be necessary to remove the effects of line emission from foreground galaxies that masquerade as Ly$\alpha$ emission.  In this section we assess the method of masking contaminated pixels from the survey to isolate Ly$\alpha$ emission.  OII (3727\AA) and H$\alpha$ (6563\AA) lines from lower redshifts will be the dominant line interlopers for Ly$\alpha$ emission for redshifts $z_{\rm Lya}>2$.  Various methods can be used to identify these interlopers and mask their corresponding pixels, including secondary line identification, photometry,and cross-correlation with templates.  These methods depend on the flux of the interloping line, in that the line must be well above a detection threshold in order to be identified.

Since it is impossible to mask all residual interloper emission, it is necessary to mask interlopers down to a tolerable residual flux such that its power spectrum is subdominant to that of Ly$\alpha$.  Using spectroscopy to identify interlopers over a large area is too time-intensive, so useful method would be to use photometric redshifts from large-area imaging survey like \emph{Euclid}.  Both instruments will have photometric redshift precisions of $\sigma_z/(1+z)\sim0.05$, allowing only the relevant populations of interlopers at the right redshifts to be masked, limiting the amount of survey area lost to masking.

One way to assess how much the flux of an interloper is decreasing as the map is being masked is a cross-correlation with another line (not Ly$\alpha$).  This cannot substitute for identifying interlopers directly for masking because the correlation between different emission line galaxy populations is not perfect enough for cleaning, but it is useful as a second check.  For $\lambda\gtrsim0.6\mu$m, we expect the intensity maps for Ly$\alpha$ to be dominated by H$\alpha$ and OII emission.  However, if we cross-correlate two maps with a wavelength ratio $\lambda_2/\lambda_1 = \lambda_{\rm H\alpha}/\lambda_{\rm OII}\sim1.8$, then we know that we are seeing the cross-correlation of H$\alpha$ and OII.  As pixels with bright H$\alpha$ emission are masked, the Ly$\alpha$ emission will begin to dominate the signal.  Considering that the auto-correlation of H$\alpha$ should be proportional to this cross-correlation, we can  use the cross-correlation as a proxy for the amount of residual H$\alpha$ remaining in the Ly$\alpha$ map.  We focus on the H$\alpha$ line because it dominates the signal over the spectral range of our hypothetical instrument (see Sec.~\ref{S:instrum}).  However, this method can only be applied for $\lambda\gtrsim0.9\mu$m because at lower wavelengths OII lines with the same redshift at H$\alpha$ lines do not appear in our spectral range.  However, this could be mitigated by adding a lower-wavelength band or partially by cross-correlating with a line higher than OII (\emph{i.e}, H$\beta$).

We estimate the difficulty of removing these interlopers by calculating the 3D power spectra of interloping H$\alpha$ and OII lines and comparing these to the Ly$\alpha$ emission power spectrum.  We determine the interloper intensities empirically using luminosity functions, specifically the H$\alpha$ luminosity function from \citet{2013arXiv1305.1399C} and the OII luminosity function from \citet{2009ApJ...701...86Z}\footnote{Although the OII luminosity function from \citet{2009ApJ...701...86Z} was determined for $z_{\rm OII}>0.75$, we extrapolate it down to $z_{\rm OII}=0.3$ as a rough estimate for our analysis.}.  Using the luminosity function, we find the unresolved luminosity density from luminosities between zero and a limiting luminosity, which can then give us the intensity, similar to Eq.~\ref{E:intj}.  Note that the limiting luminosity $L_{\rm lim}$ depends on the redshift of the interloper and the limiting flux $f_{\rm lim}$ of the detector with these quantities being related by $L_{\rm lim}=4\pi d_L^2(z)f_{\rm lim}$.  We also include wavevector translation effects in the interloper power spectrum since the contribution of interlopers to the number density of sources will be counted per 3D pixels at the Ly$\alpha$ redshift, not the interloper redshift.  To account for small scales translating to larger scales, we include nonlinear clustering in the power spectrum according to the halo model \citep{2002PhR...372....1C}.  More details on the formalism for LSS surveys will be presented in an upcoming paper [Pullen et al.~(2013)].

\begin{figure}
\begin{center}
{\scalebox{.5}{\includegraphics[width=\textwidth]{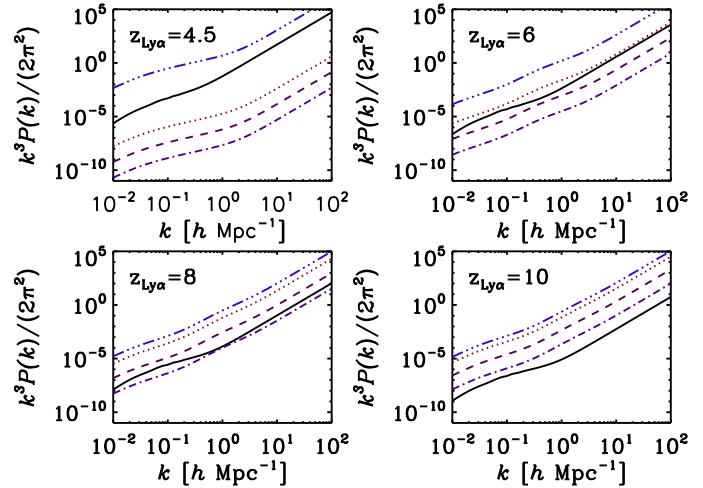}}}
\caption{\label{F:clintHa} The 3D auto-power spectra for Ly$\alpha$ (solid line), along with the corresponding H$\alpha$ interloper power spectra (dot-dot-dot-dashed) and residual spectra with $f_{\rm lim}=10^{-16}$ (dotted), $10^{-17}$ (short-dashed), and $10^{-18}$ (dot-dashed) erg s$^{-1}$ cm$^{-2}$.  The unit for the $y$-axis is nW$^2$ m$^{-4}$ sr$^{-2}$.}
\end{center}
\end{figure}

\begin{figure}
\begin{minipage}[b]{0.48\linewidth}
\centering
\includegraphics[width=\textwidth]{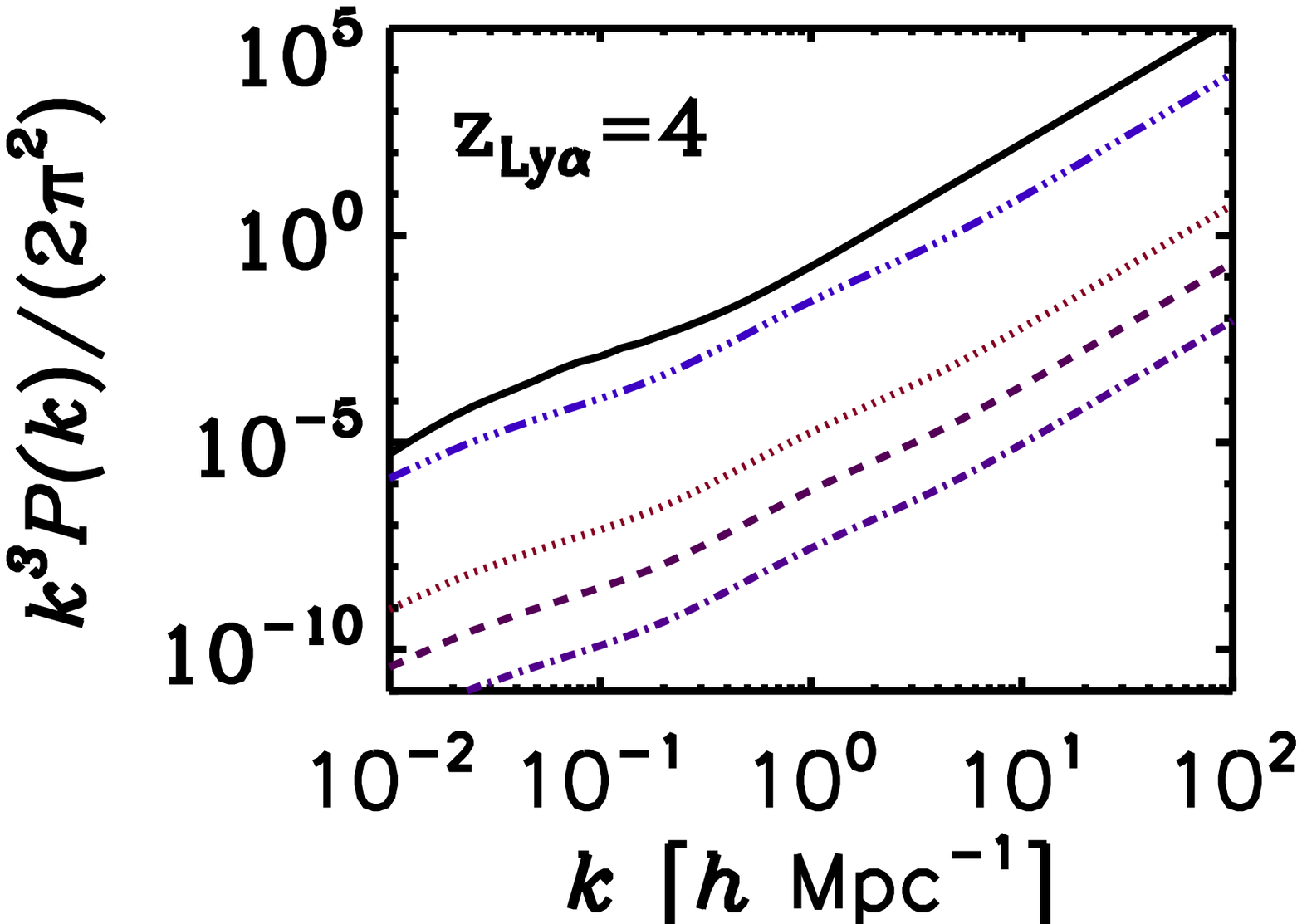}
\end{minipage}
\hspace{0.1cm}
\begin{minipage}[b]{0.48\linewidth}
\centering
\includegraphics[width=\textwidth]{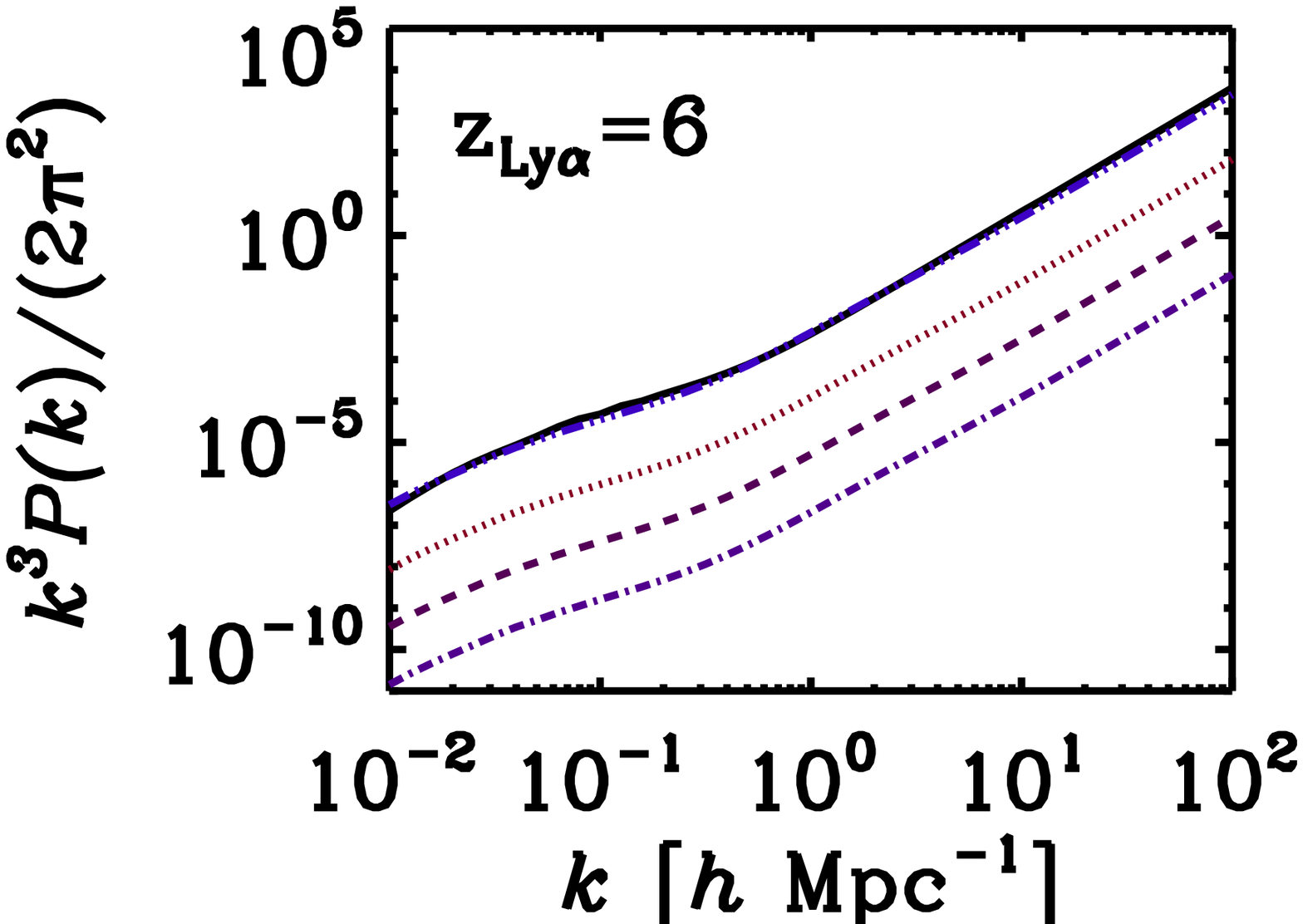}
\end{minipage}
\begin{center}
\caption{\label{F:clintO2} The 3D auto-power spectra for Ly$\alpha$ (solid line), along with the corresponding OII interloper power spectra (dot-dot-dot-dashed) and residual spectra with $f_{\rm lim}=10^{-16}$ (dotted), $10^{-17}$ (dashed), and $10^{-18}$ (dot-dashed) erg s$^{-1}$ cm$^{-2}$.  The unit for the $y$-axis is nW$^2$ m$^{-4}$ sr$^{-2}$.}
\end{center}
\end{figure}

We plot the Ly$\alpha$ power spectrum along with the H$\alpha$ and OII interloper power spectra in Figs.~\ref{F:clintHa} and \ref{F:clintO2}.  Note that we can only estimate the OII interloper emission up to $z_{\rm Ly\alpha}=6$ because the luminosity function is only estimated up to $z_{\rm OII}=1.45$.  From these plots, we see that measuring the Ly$\alpha$ power spectra at low redshifts ($z\lesssim5$) requires eliminating H$\alpha$ down to $f_{\rm lim}=10^{-16}$ erg s$^{-1}$ cm$^{-2}$, while $z_{\rm Ly\alpha}\sim6$ would require eliminating H$\alpha$ down to $f_{\rm lim}=10^{-17}$ erg s$^{-1}$ cm$^{-2}$ for $k>1h/$Mpc and lower for larger scales.  At $z_{\rm Ly\alpha}=6$, all the signal in our instrumental configuration is from smaller scales, so the $f_{\rm lim}=10^{-17}$ erg s$^{-1}$ cm$^{-2}$ cut is sufficient.  We now estimate what AB magnitude in an optical band would correspond to our required cut in H$\alpha$.  Using the COSMOS mock catalog \citep{2011A&A...532A..25J}, we find that for $F_{\rm H\alpha}=10^{-17}\,{\rm erg\,s^{-1}cm^{-2}}$, the required cut in the $r$ band is $m_r\sim26.5$, which is close to the sensitivity of upcoming surveys over hundreds of square degrees (e.g. HSC).  For higher redshifts, the required sensitivity to H$\alpha$ emitters will be progressively lower, even as low as $f_{\rm lim}=10^{-20}$ erg s$^{-1}$ cm$^{-2}$ for $z_{\rm Ly\alpha}=10$ which is not expected for the foreseeable future, making this type of cleaning not feasible for high redshifts.  We also see that OII without masking is already comparable to the Ly$\alpha$ signal.  Thus, the OII maps should not be too difficult to clean.

\begin{figure}
\begin{center}
{\scalebox{.5}{\includegraphics[width=\textwidth]{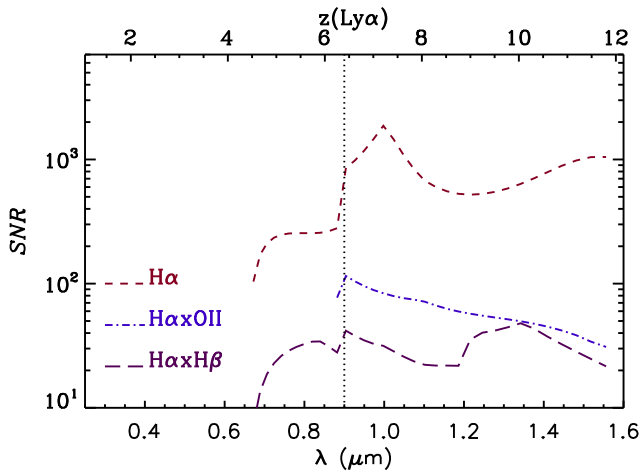}}}
\caption{\label{F:hao2snrz} Forecasts for SNR of the H$\alpha$, H$\alpha\times$OII, and H$\alpha\times$H$\beta$ 3D power spectra in spectral bins for various instrumental configurations.  The spectral bin widths for the lines are determined by the spectral resolution, which is $R=40$.  The bars denote the SNR over redshift ranges $\Delta z_{\rm Ly\alpha}=1$.  The dotted, vertical line divides the first and second spectral bands.  We see that all three power spectra should be detectable before masking.}
\end{center}
\end{figure}

In Fig.~\ref{F:hao2snrz} we plot both the H$\alpha$, H$\alpha\times$OII (for $\lambda>0.9\mu$m), and H$\alpha\times$H$\beta$ power spectra for our hypothetical instrument without masking to gauge their utility as a proxy for H$\alpha$ contamination.  As with the Ly$\alpha$ $SNR$, we limit $k_{\rm min}$ by the volume of the H$\alpha$ survey.  To estimate the H$\alpha\times$H$\beta$ cross-power, we use the OII luminosity function from \citet{2009ApJ...701...86Z} and assume the line ratio H$\alpha$/H$\beta$ = 0.28 \citep{2011A&A...532A..25J}.  Note that for $z_{\rm Ly\alpha}<4.5$, H$\alpha$ is no longer an interloper, and OII should be easy to remove.  We see that the H$\alpha$ power spectra has a high SNR.  Note that the Ly$\alpha$ redshift bins correspond to $\Delta z_{\rm H\alpha}/(1+z_{\rm H\alpha})=1/R=0.025$, which is about half as small as the expected photometric redshift errors in upcoming surveys, possibly requiring some of the maps to be combined.  For the cross-correlations, both the H$\alpha\times$OII and H$\alpha\times$H$\beta$ signals have detectable SNRs.  This result implies that these cross-correlations can be used as a proxy for H$\alpha$ contamination.

Now that we know the required $f_{\rm lim}$ for a low-redshift Ly$\alpha$ measurement, we can estimate for our instrument the sky fraction that must be masked based on the pixel size.  This will affect the sky fraction of the survey and thus the SNR of the angular power spectrum.  It can be shown that the fraction of the survey with pixel size $\Omega_{\rm pix}$ that must be removed to cut an areal number density $n_{\rm int}$ of interlopers is given by $f_{\rm mask}=n_{\rm int}\Omega_{\rm pix}$ (for $f_{\rm mask}<1$).  Note that with the volume density $N(<L_{\rm lim},z)=\Phi_*(z)\Gamma[\alpha(z)+1,L_{\rm lim}/L_*(z)]$, the areal density $n_{\rm int}$ is written as
\begin{eqnarray}
n_{\rm int}\simeq\Delta z\frac{d\chi}{dz}\chi(z)^2N(<L_{\rm lim},z)\, ,
\end{eqnarray}
where $\Delta z =(1+z)/R$.  For $f_{\rm lim}=10^{-14}$ erg s$^{-1}$ cm$^{-2}$ at $z_{\rm Ly\alpha}=4.5$ ($z_{\rm H\alpha}=0.02$), the residual H$\alpha$ population has a number density of 4.2$\times10^{-4}$ arcmin$^{-2}$, implying the required fraction of the survey that must be masked is completely negligible.  For $z_{\rm Ly\alpha}=6$, where $f_{\rm lim}=10^{-17}$ erg s$^{-1}$ cm$^{-2}$ is needed, $n_{\rm int}=0.15$ arcmin$^{-2}$, implying $f_{\rm mask}=0.16$\%, which will still be a very small effect.  Thus, we expect to be able to remove line interlopers from redshifts $z_{\rm Ly\alpha}\lesssim7$ without degrading the signal.  Even at $z_{\rm Ly\alpha}=8$, which requires $f_{\rm lim}\sim10^{-19}$ erg s$^{-1}$ cm$^{-2}$, $n_{\rm int}=1.6$ arcmin$^{-2}$ and the masked sky fraction is only $f_{\rm mask}=1.75$\%.

\section{Conclusions} \label{S:conc}
We have constructed a model of the minimum Ly$\alpha$ line emission and perturbations over cosmic time, between the redshifts $z=2-12$, that satisfy the constraint of sustaining reionization.  This model was constructed to assess the viability of a Ly$\alpha$ intensity mapping survey both to constrain cosmological parameters and to probe the reionization epoch and star formation at low and high redshifts.  For star-forming, massive dark-matter halos, or Ly$\alpha$ emitters, various sources of emission were considered.  Halo emission was the dominant source for Ly$\alpha$ emitters over all redshifts.  This signal was also found to be consistent with empirical estimates based on measured luminosity functions.  In addition, we also considered line emission from the diffuse IGM, consisting of a recombinations component and a stellar continuum component.  The Ly$\alpha$ intensity signal was found to be dominated by diffuse IGM emission at all redshifts $z\lesssim10$.

The 3D power spectrum from Ly$\alpha$ line intensity was also found to be dominated by  diffuse IGM emission.  A sensitive spectrometer can allow significant measurements of low- and mid-redshift Ly$\alpha$ fluctuations through intensity mapping, but EoR fluctuations appear to be challenging to measure by these studies.  The detectability of the fluctuations signal requires masking of line emission from low-redshift galaxies, using spectral and photometric methods.

It should be noted that our model is very dependent on our fiducial parameters, particularly at high redshifts.  The signal at high redshifts was set to the minimum star formation rate and the minimum required to sustain reionization, and it is conceivable for the power spectrum signal at these redshifts to increase by a factor of 100, placing it within a detectable range.  A measurement of Ly$\alpha$ intensity fluctuations at high significance could allow probes of large scale structure, star formation, and reionization over cosmic history.

\begin{acknowledgments}

We thank P.~Capak, A.~Cooray, S.~Furlanetto, Y.~Gong, C.~Hirata, A.~Lidz, M.~Viero, and M.~Zemcov for helpful comments and useful discussions. Part of the research described in this paper was carried out at the Jet Propulsion Laboratory, California Institute of Technology, under a contract with the National Aeronautics and Space Administration. AP was supported by an appointment to the NASA Postdoctoral Program at the Jet Propulsion Laboratory, California Institute of Technology, administered by Oak Ridge Associated Universities through a contract with NASA. This work was supported by the Keck Institute of Space Studies and we thank colleagues at the ``First Billion Years'' for stimulating discussions, in particular J. Bowman and A. Readhead for organizing it.
\end{acknowledgments}

\end{document}